\documentclass[aps,prl,reprint,superscriptaddress]{revtex4-1}

\usepackage[colorlinks,linkcolor=red,urlcolor=blue,citecolor=blue]{hyperref}
\usepackage{relsize}
\usepackage{graphicx}
\usepackage{amsmath, bbm}
\usepackage{amssymb}
\usepackage{amsthm}
\usepackage{mathrsfs}
\usepackage{xcolor}
\usepackage{cancel}
\usepackage{titlesec}
\usepackage{enumitem,braket}  
\usepackage{tikz}  
\usetikzlibrary{arrows,shapes,positioning,shadows,backgrounds,fit}
\usepackage[normalem]{ulem}
\titlespacing{\section}{0ex}{2ex}{0.4ex}
\titleformat{name=\section}{\bf}{\thesection.}{.3em}{}
\usepackage{soul}
\usepackage{dsfont}

\def\be{\begin{eqnarray}}
\def\ee{\end{eqnarray}}
\newcommand{\tr}[1]{\text{Tr}\left(#1\right)}
\newcommand{\Tr}[1]{\text{tr}\left(#1\right)}

\renewcommand{\ln}[1]{\mathrm{ln} \left({#1} \right)}

\newcommand{\bh}{\mathcal{B}(\mathcal{H})}

\newcommand{\id}{\hat{\mathbb{I}}}

\newcommand{\lind}{\mathscr{L}}

\newcommand{\com}{{\mathcal{C}}}

\newcommand{\tg}{{\mathcal{T}}}

\begin{document}

\title{Quantum fluctuations hinder finite-time information erasure near the Landauer limit}

\author{Harry J.~D. Miller}
\affiliation{Department of Physics and Astronomy, The University of Manchester, Manchester M13 9PL, UK}

\author{Giacomo Guarnieri}
\affiliation{School of Physics, Trinity College Dublin, College Green, Dublin 2, Ireland}

\author{Mark T.~Mitchison}
\affiliation{School of Physics, Trinity College Dublin, College Green, Dublin 2, Ireland}

\author{John Goold}
\email{gooldj@tcd.ie}
\affiliation{School of Physics, Trinity College Dublin, College Green, Dublin 2, Ireland}

\date{\today}

\begin{abstract}
Information is physical but information is also processed in finite time. Where computing protocols are concerned, finite-time processing in the quantum regime can dynamically generate coherence. Here we show that this can have significant thermodynamic implications. We demonstrate that quantum coherence generated in the energy eigenbasis of a system undergoing a finite-time information erasure protocol yields rare events with extreme dissipation. These fluctuations are of purely quantum origin. By studying the full statistics of the dissipated heat in the slow driving limit, we prove that coherence provides a non-negative contribution to all statistical cumulants. Using the simple and paradigmatic example of single bit erasure, we show that these extreme dissipation events yield distinct, experimentally distinguishable signatures. 
\end{abstract}

\maketitle

Landauer's principle states that any logically irreversible computation produces entropy, which dissipates heat to non-information bearing degrees of freedom~\cite{landauer1961irreversibility}. This basic principle not only sets an ultimate physical limit to information processing but also forms the foundation of the thermodynamics of computation~\cite{bennett1982thermodynamics,lloyd2000ultimate} and information~\cite{sagawa2012thermodynamics,parrondo2015thermodynamics,goold2016role,vinjanampathy2016quantum,binder2018thermodynamics}, while playing a pivotal role in the resolution of the Maxwell demon paradox~\cite{plenio2001physics,Maruyama}. The most elementary logically irreversible process is the erasure of one bit of information, which dissipates an amount of heat $q \geq k_{B}T\ln {2}$ to the environment, where $k_{B}$ is Boltzmann's constant and $T$ is the temperature. This fundamental lower bound on dissipated heat is known as the Landauer limit.

In reality, any physical implementation of information erasure takes place under non-equilibrium conditions, where a possibly microscopic system (information bearing degree of freedom) is manipulated in finite time while in contact with a heat bath. In this setting, fluctuations become significant and path-dependent thermodynamic quantities, such as heat and work, are described by probability distributions~\cite{sekimoto2010,jarzynski2011,seifert2012,esposito2009,campisi2011,hanggi2015}. This has important consequences for heat management in nanoscale devices, which must be designed to tolerate large and potentially destructive fluctuations. As information processing technology encroaches on the small scale where quantum effects take hold, it thus becomes crucial to understand how quantum as well as thermal fluctuations contribute to dissipation during the erasure process. 
 
Minimising dissipation typically requires slow driving in order to remain in the quasi-static regime. This has been highlighted by the first generation of experiments aiming to experimentally study information erasure near the Landauer limit on both classical~\cite{orlov2012experimental,Arakelyan2012,jun2014,roldan2014universal,hong2016experimental,Gavrilov2016} and quantum~\cite{peterson2016experimental,yan2018single,gaudenzi2018quantum,saira2020nonequilibrium} platforms. In particular, the probability distributions of work and heat during a finite-time protocol were extracted in pioneering experiments on Brownian particles confined by tunable double-well potentials~\cite{Arakelyan2012,jun2014}. In the quasi-static regime, it was found that the dissipated heat approaches the Landauer limit on average~\cite{Arakelyan2012}, while its probability distribution becomes Gaussian~\cite{jun2014} with a variance constrained by the work fluctuation-dissipation relation~\cite{jar}. However, experiments exploring the full heat statistics of quasi-static erasure have so far been limited to a classical regime, leaving open the question of how quantum effects influence the heat distribution.

\begin{figure}
    \centering
    \includegraphics[width=0.8\columnwidth]{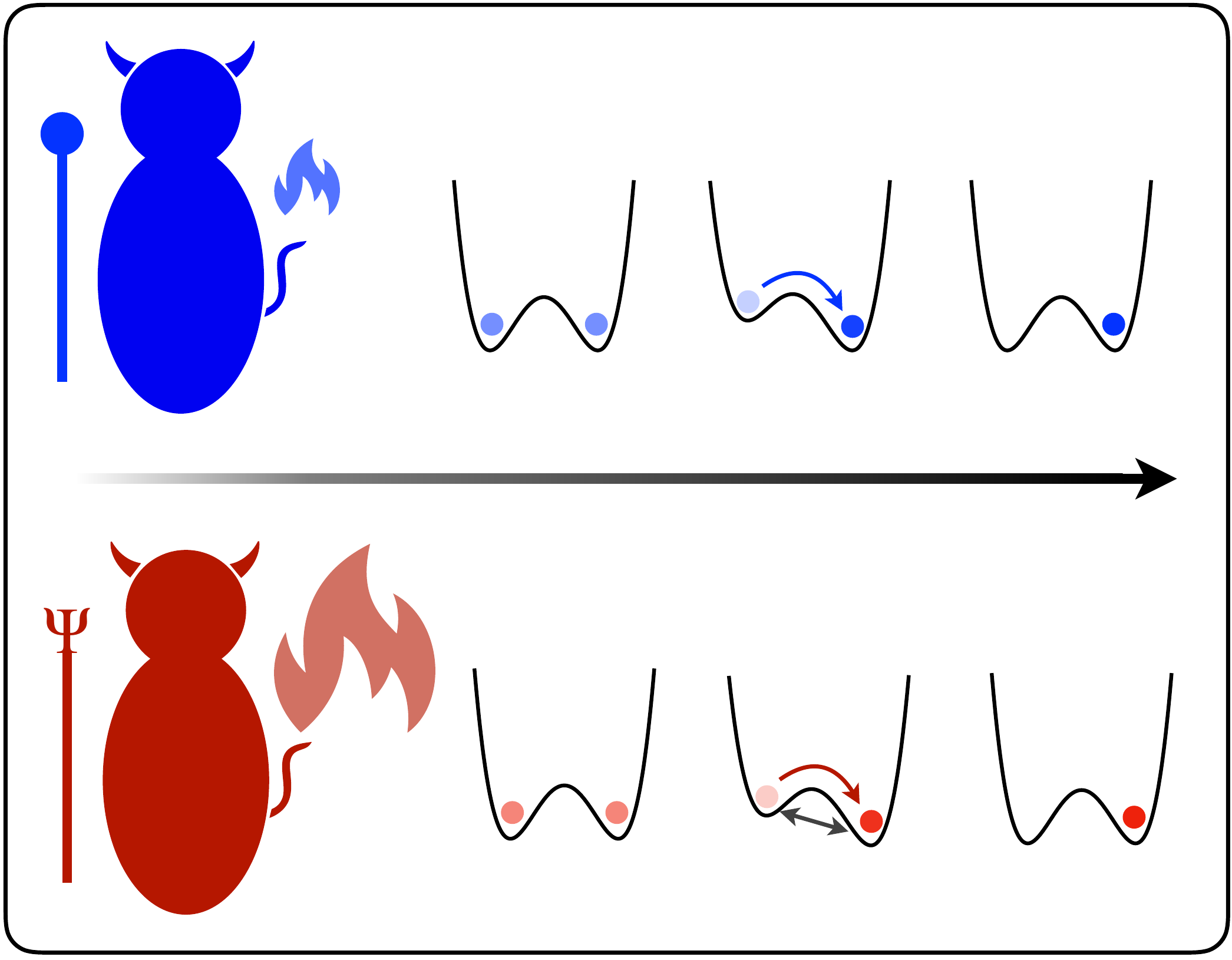}
    \caption{Schematic of the erasure protocol and our main results. An intelligent being (demon) performs erasure through a controlled process that resets a physical bit of information to a fixed reference state. In this example, the bit is encoded in the position of a particle confined by a double-well potential. Classically, erasure is performed by raising the potential of one well until thermal fluctuations drive the particle into the lower-energy state, at the cost of dissipating some heat into the environment. Quantum mechanics allows the particle to coherently tunnel under the barrier as well as hop over it, leading to large quantum fluctuations in the dissipated heat.\label{fig:schematic}}
\end{figure}

Here we demonstrate that quantum coherence is always detrimental for the attainability of the fundamental Landauer limit during slow erasure protocols. More precisely, we prove that quantum coherence generated in the energy eigenbasis of a slowly driven system yields a non-negative contribution to all statistical cumulants of the dissipated heat and renders the associated probability distribution non-Gaussian. Coherent control therefore increases the overall likelihood of dissipation above the Landauer bound due to the heat distribution developing a significant skewness. We exemplify this general principle by studying the erasure of one bit of information stored in a quantum two-level system, as illustrated schematically in Fig.~\ref{fig:schematic}. We find that quantum fluctuations generate distinct and, in principle, experimentally distinguishable signatures in the heat statistics, consisting of rare events with extreme dissipation $q\gg k_BT$. Despite their rarity, the significance of such processes is clear in light of the many billions of bits that are irreversibly processed each second in modern computer hardware. Aside from unambiguously demonstrating a quantum effect in information thermodynamics, our findings imply that control strategies designed to suppress quantum fluctuations may be necessary to mitigate dissipation in miniaturised information processors, in agreement with results from single-shot statistical mechanics~\cite{Browne2014PRL}.
 
 {\bf Erasure protocol.} We note that Landauer's principle for {\it finite} quantum baths~\cite{esposito2010entropy,reeb2014improved,goold2014measuring,goold2015nonequilibrium,timpanaro2019landauers} has recently been experimentally explored in \cite{peterson2016experimental,yan2018single}. In this work, we consider an erasure protocol where a controllable quantum system with encoded information is continuously connected to non-information bearing degrees of freedom modelled as an {\it infinite} heat bath. Specifically, information encoded in a quantum system of finite dimension $d$, described by a maximally mixed state $\id/d$, is erased by bringing the system to its ground state $\ket{0}\bra{0}$, resulting in a decrease in information entropy $\Delta S=-\log d$. This is achieved by slowly varying a control Hamiltonian $\hat{H}_t$ over a finite time interval $t\in[0,\tau]$ while the system is weakly coupled to a thermal reservoir at inverse temperature $\beta = 1/k_BT$. We assume Markovian dynamics generated by an adiabatic Lindblad equation~\cite{Albash2012}, $\dot{\hat{\rho}}_t=\mathscr{L}_t(\hat{\rho}_t)$, where the generator $\mathscr{L}_t$ obeys quantum detailed balance with respect to the Hamiltonian $\hat{H}_t$ at all times~\cite{Alicki1976}. This condition ensures a thermal instantaneous fixed point, $\mathscr{L}_t(\hat{\pi}_t)=0$, where $\hat{\pi}_t=e^{-\beta \hat{H}_t}/\tr{e^{-\beta \hat{H}_t}}$. Erasure can be realised by first taking an initial Hamiltonian with $\hat{H}_0\simeq 0$ relative to the thermal energy $k_BT$, then increasing its energy gaps until they far exceed $k_BT$. If one assumes that the system is in equilibrium at the end of the process, this results in effective boundary conditions $\hat{\rho}_0=\hat{\pi}_0\simeq \id/d$ and $\hat{\rho}_\tau=\hat{\pi}_\tau\simeq \ket{0}\bra{0}$. 
 
{\bf Heat statistics.} Having introduced our erasure protocol, we now discuss the full counting statistics of the dissipated heat. In the weak-coupling limit, heat is unambiguously identified with the change in energy of the reservoir~\cite{hangnew}. For Lindblad dynamics with detailed balance, the evolution may be unravelled into quantum-jump trajectories~\cite{Plenio1998}, where heat exchange is associated with the emission and absorption of energy quanta by the driven quantum system ~\cite{Horowitz2012,Horowitz2014,Manzano2015,Manzano2018}. Operationally, each trajectory represents an individual run of an experiment in which the environment is continuously monitored by direct detection of the emitted and absorbed quanta~\cite{Pekola2013b}. This is formally described by a set of coarse-grained time points at which measurements occur, $\tau\geq t_N\geq \ldots \geq t_1 \geq t_0 = 0$, separated by an increment $\delta t$ much smaller than the characteristic timescale of dissipation. The system evolution from time $t_n\to t_{n+1}$ is given by the quantum channel $\tg_{n}:= e^{\delta t \lind_{t_n}} = \sum_{x_n} \tg_{x_n}$, where $\tg_{x_n}(\cdot) = \hat{K}_{x_n}(t_n) (\cdot) \hat{K}_{x_n}^\dagger(t_n)$ form a set of Kraus operators satisfying $\sum_{x_n}\hat{K}_{x_n}^\dagger(t_n)\hat{K}_{x_n}(t_n)=\id$ and $x_n$ labels the distinguishable outputs of the detector. Each trajectory of the open system is then specified by its measurement record, i.e.\ a sequence of the form $\Gamma := \{x_0, \dots , x_N\}$ occurring with probability 
\begin{align}\label{eq:trajecprob}
p(\Gamma) := \frac{1}{d} \bra{0} \prod^{N}_{n=0} \tg_{x_n}( \id) \ket{0}.
\end{align}
To ensure detailed balance~\cite{Fagnola2007,Manzano2015}, the Kraus operators are taken to satisfy $[\hat{H}_t,\hat{K}_{x}(t)] = -\hbar\omega_x(t)\hat{K}_x(t)$, where $\hbar\omega_x(t)$ are differences between the eigenvalues of $\hat{H}_t$. Thus, $\omega_x>0$ ($\omega_x<0)$ represents a detected emission (absorption) while $\omega_x=0$ represents no detection. This assumption ensures that heat entering the environment may be  identified along each trajectory $\Gamma$, being given by the sum of these energy changes:
\begin{align}\label{eq:heat}
q(\Gamma):=-\sum^N_{n=0}\hbar \omega_{x_n}(t_n).
\end{align}
We note that in the weak-coupling regime this is equivalent to the outcome of a two-point measurement of the environment's energy at the beginning and end of the protocol~\cite{Roeck2007,Silaev2014a,Liu2016a}. The average heat flux is given by $\langle \dot{q} \rangle=\tr{\hat{H}_t \dot{\hat{\rho}}_t}$, consistent with well known results for weak-coupling Lindblad dynamics \cite{Alicki1979}. 

It is convenient to define the excess stochastic heat
\begin{align}
\label{excess_heat}
    \tilde{q}(\Gamma):=q(\Gamma)-k_BT \log d,
\end{align}
which quantifies the additional heat in excess of the Landauer bound. The full statistics of excess heat can be obtained from the cumulant generating function (CGF), evaluated in the continuum limit $\delta t\to 0$:
\begin{align}\label{eq:CGF}
\mathcal{K}_q(u):=\text{ln} \ \sum_{\{\Gamma \}} e^{- u \tilde{q}(\Gamma)} p(\Gamma).
\end{align} 
This provides the cumulants according to $\kappa_k=(-1)^k\frac{d^k}{du^k}\mathcal{K}_q(u)\big|_{u=0}$, where $\kappa_1=\langle q \rangle- k_BT \log d $ is the average excess heat, $\kappa_2=\text{Var}(q)=\langle q^2\rangle-\langle q \rangle^2$ is the variance, and so forth.

{\bf The role of coherence in erasure.} We now come to the main finding of our work, namely that quantum coherence generates additional dissipation during information erasure. We focus on protocols close to the quasi-static limit, where the dissipation approaches the Landauer bound. This requires the Hamiltonian to be driven slowly relative to the relaxation timescale of the dynamics, implying that the system state remains close to equilibrium at all times. We may therefore use an expansion of the form $\hat{\rho}_t \simeq \hat{\pi}_t + \tau^{-1} \delta\hat{\rho}_t$, with $\delta\hat{\rho}_t$ a linear-order perturbation to the equilibrium state, $\hat{\pi}_t$. 

Neglecting corrections of order $\mathcal{O}(\tau^{-2})$, we find that the full statistics of excess heat in the slow-driving limit can be separated into a classical and quantum part (see Appendix \ref{app:split}):
\begin{align}\label{eq:split}
    \mathcal{K}_q(u)=\mathcal{K}^d_q(u)+\mathcal{K}^c_q(u).
\end{align}
Due to the additivity of the CGFs we may interpret the total excess heat as a sum of two independent random variables, $\tilde{q}(\Gamma)=\tilde{q}_d(\Gamma)+\tilde{q}_c(\Gamma)$, with $\tilde{q}_d(\Gamma)$ described by a classical (\textit{d}iagonal) CGF $\mathcal{K}^d_q(u)$ and cumulants of $\tilde{q}_c(\Gamma)$ given by the quantum (\textit{c}oherent) CGF $\mathcal{K}^c_q(u)$. These different contributions to the heat statistics relate directly to the different ways a quantum state can evolve, through changes to either the populations or the coherences in the energy eigenbasis (see Appendix \ref{app:Renyi}). Specifically, the classical CGF is given by
\begin{align}\label{eq:CGFd_renyi}
    \mathcal{K}^d_q(u)=  k_BT(u-k_BTu^2)\int^\tau_0 dt \ \frac{\partial}{\partial t}S(\mathcal{D}_{\hat{H}_s}(\hat{\rho}_t) || \hat{\pi}_{s})\bigg|_{s=t},
\end{align}
where $S(\hat{\rho} || \hat{\sigma})=\tr{\hat{\rho}  \ \ln{\hat{\rho}}}-\tr{\hat{\rho} \ \ln{\hat{\sigma}}}$ is the quantum relative entropy and ${\mathcal{D}_{\hat{H}_t}(.)=\sum_{n}\ket{n_t}\bra{n_t}(.) \ket{n_t}\bra{n_t}}$ denotes the dephasing map in the {\it instantaneous} energy eigenbasis $\{ \ket{n_t}\}$ of $\hat{H}_t$. Eq.~\eqref{eq:CGFd_renyi} expresses the fact that classical contributions to the excess heat occur when the system populations deviate from the instantaneous Boltzmann distribution. Furthermore, the quantum CGF can be identified as
\begin{align}\label{eq:CGFq_renyi}
    \mathcal{K}^c_q(u) = -u k_BT \int^\tau_0 dt \ \frac{\partial}{\partial t}S_{1-u k_B T}\big(\hat{\rho}_t || \mathcal{D}_{\hat{H}_s}(\hat{\rho}_t)\big)\bigg|_{s=t},
\end{align}
where $S_\alpha(\hat{\rho}||\hat{\sigma})=(\alpha-1)^{-1}\mathrm{ln}\tr{\hat{\rho}^\alpha\hat{\sigma}^{1-\alpha}}$ for $\alpha\in(0,1)\cup(1,\infty)$ represents the quantum Renyi divergence. The function $S_{\alpha}\big(\hat{\rho}_t || \mathcal{D}_{\hat{H}_t}(\hat{\rho}_t)\big)$ can be interpreted as a proper measure of asymmetry with respect to the instantaneous energy eigenbasis \cite{Lostaglio2015d}, which is closely related to the amount of coherence contained in the state~\cite{Streltsov2017}. The first cumulant of Eq.~\eqref{eq:CGFq_renyi} is proportional to the relative entropy of coherence \cite{Baumgratz2014}, which has previously been identified as a quantum contribution to average entropy production in open~\cite{Santos2019,Mohammady,Mohammady2020a} and closed~\cite{Francica2017} systems. A similar division into classical and quantum components was obtained in Ref.~\cite{Scandi2020} for work statistics in the slow-driving limit. 

Remarkably, the splitting embodied by Eq.~\eqref{eq:split} puts constraints of non-classical origin on the full statistics of dissipated heat. To see this, let us first convert the diagonal part $\mathcal{K}^d_q(u)$ into a probability distribution via an inverse Laplace transform. This yields a Gaussian distribution with mean and variance connected by $\langle \tilde{q}_d \rangle=\frac{1}{2}\beta \text{Var}(\tilde{q}_d)$, as expected for a classical process in the slow-driving limit \cite{Speck}. It follows that the classical heat distribution obeys the Landauer bound, $\langle \tilde{q}_d \rangle\geq 0$. Turning to the quantum contribution, no such straightforward expression can be obtained for the distribution $P(\tilde{q}_c)$ due to the complicated dependence of the quantum covariance~\eqref{eq:CGFq_renyi} on the counting field $u$. Despite this, one may prove that the cumulants of $\tilde{q}_c$ are all monotonically non-decreasing in time (see Appendix \ref{app:mono}):  \begin{align}\label{eq:mono}
    (-1)^k\frac{d^k}{du^k}\dot{\mathcal{K}}^c_q(u)\bigg|_{u=0}\geq 0, \ \ \ \ \forall k.
\end{align}
This immediately implies that coherence imparts a non-negative contribution to the mean heat dissipated during erasure, i.e.
\begin{align}
    \langle q \rangle= k_B T \ln{d}+ \langle \tilde{q}_d \rangle+ \langle \tilde{q}_c \rangle, \ \ \ \text{with} \ \ \ \langle \tilde{q}_c \rangle\geq 0.
\end{align}
Furthermore, all higher cumulants are also non-negative, implying increased fluctuations that will generally exhibit positive skew and kurtosis. As a consequence, the overall heat distribution can be highly non-Gaussian, in stark contrast to the classical case. 

These results have profound repercussions for the erasure of information stored in a quantum system. Manipulating such a system in finite time typically generates coherence due to the presence of several non-commuting terms in the Hamiltonian, a feature which is unavoidable for certain physical architectures. Not only does this lead to a greater energetic cost on average, it also increases the probability of large fluctuations where a quantity of heat $q \gg k_B T\ln{d}$ well above the Landauer bound is dissipated into the surroundings. 

{\bf Example: qubit erasure.} To illustrate our findings, we consider an elementary example of erasure where information is stored in a quantum two-level system described by the Hamiltonian
\begin{equation}
    \label{qubit_Hamiltonian}
    \hat{H}_t = \frac{\varepsilon_t}{2}\left( \cos\theta_t\hat{\sigma}_z + \sin\theta_t \hat{\sigma}_x\right).
\end{equation}
This generic Hamiltonian describes the low-energy dynamics of a particle in a double-well potential~\cite{Leggett1987} or a genuinely discrete information storage device such as a charge or spin qubit~\cite{Hanson2007}. Thermal dissipation is modelled by a bosonic heat bath described by an adiabatic Lindblad master equation in the limit of slow driving and weak coupling~\cite{Albash2012}; see Appendix~\ref{app:example} for details. Stored information is erased by increasing the energy splitting $\varepsilon_t$ from its initial value, $\varepsilon_0\approx 0$, to a final value, $\varepsilon_\tau\gg k_BT$, leaving the qubit in its ground state with near-unit probability. The mixing angle $\theta_t$ encapsulates the competition between energetic bias $(\hat{\sigma}_z)$ and coherent tunnelling $(\hat{\sigma}_x)$. If $\theta_t$ is constant, Eq.~\eqref{qubit_Hamiltonian} describes a classical bit. Conversely, when $\dot{\theta}_t \neq 0$ --- which will generally be the case, e.g.~for quantum double-well systems --- the protocol is non-commuting.

\begin{figure}
\flushleft \hspace{0.1cm} {(a)} \hspace{3.3cm} {(b)}\\
    \centering
    \includegraphics[width=\linewidth, trim = 20mm 165mm 20mm 175mm, clip]{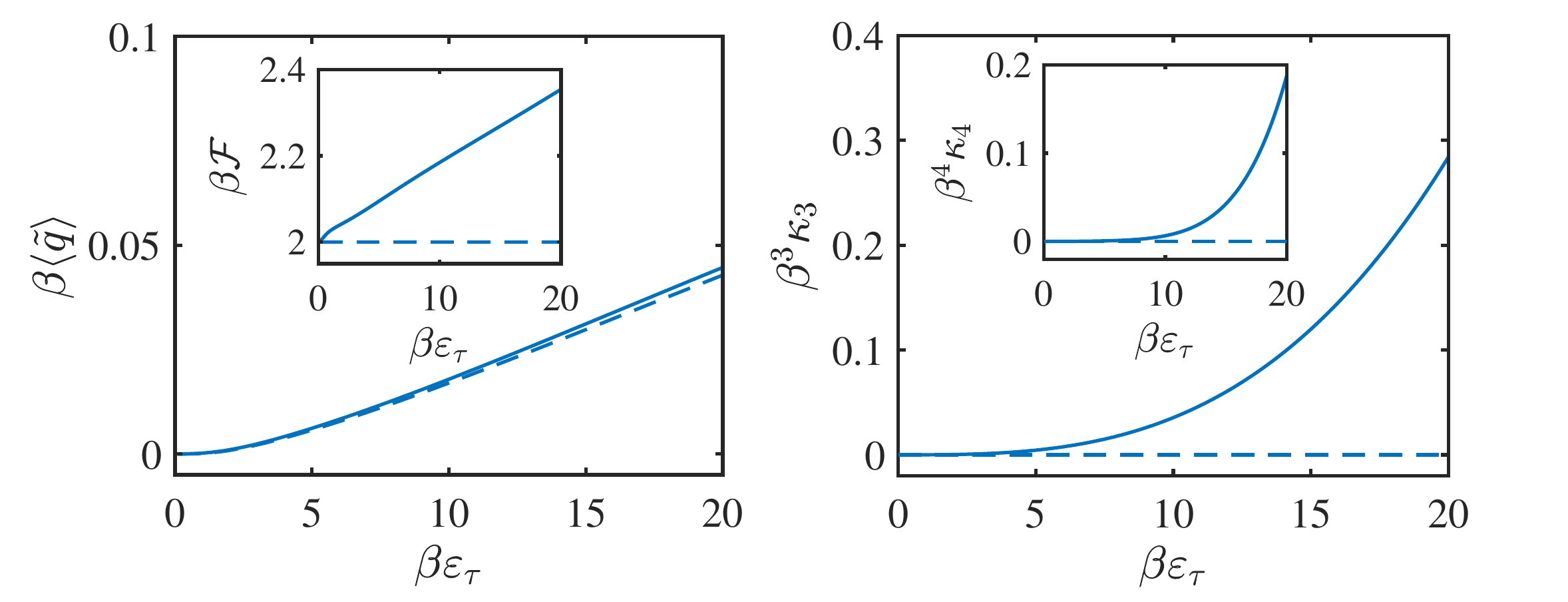}
    \caption{Heat statistics of slow driving processes. (a) Mean excess heat (main) and Fano factor, $\mathcal{F}$ (inset). (b) Third cumulant (main) and fourth cumulant (inset) of the heat distribution, demonstrating non-Gaussian statistics. Solid lines show a quantum protocol with ${\varepsilon_t = \varepsilon_0 + (\varepsilon_\tau- \varepsilon_0)\sin^2(\pi t/2\tau)}$ and ${\theta_t = \pi(t/\tau-1)}$, dashed lines show the corresponding classical protocol with identical $\varepsilon_t$ but $\theta_t=0$. The initial energy splitting is $\varepsilon_0 = 0.02\varepsilon_\tau$ and the protocol duration is $\tau = 250/\bar{\gamma}$, where $\bar{\gamma}$ is a characteristic thermalisation rate given by the time average of $\gamma_t = \tfrac{1}{2}\hbar^{-1}\alpha \varepsilon_t \coth(\beta\varepsilon_t/2)$, with $\alpha=0.191$ the coupling to an Ohmic bath.  \label{fig:cumulants}}
\end{figure}

In Fig.~\ref{fig:cumulants} we plot the first four cumulants of the heat distribution, comparing a quantum protocol to a classical process with identical $\varepsilon_t$ but $\dot{\theta}_t=0$. These analytical results are derived in the slow-driving limit at order $\mathcal{O}(\tau^{-1})$ (see Appendix~\ref{app:example}). We show in Fig.~\ref{fig:cumulants}(a) that the mean excess heat [Eq.~\eqref{excess_heat}] takes small but non-zero values in the erasure regime, $\varepsilon_\tau\gg k_BT$, reflecting the entropy produced in this finite-time process. While the quantum and classical protocols show similar dissipation on average, they differ significantly in their fluctuations. The inset of Fig.~\ref{fig:cumulants} shows the Fano factor, $\mathcal{F}={\rm Var}(\tilde{q})/\langle \tilde{q}\rangle$, which is increased by quantum fluctuations above the classical value, $\mathcal{F} = 2 k_B T + \mathcal{O}(\tau^{-2})$, that follows from the fluctuation-dissipation relation. The most significant difference arises in higher-order statistics: the non-classical nature of the heat distribution is witnessed by its third and fourth cumulants, shown in Fig.~\ref{fig:cumulants}(b). These imply significant skewness and kurtosis and signal the presence of non-Gaussian tails in the distribution. 

\begin{figure}
    \centering
    \includegraphics[width=\linewidth, trim = 5mm 75mm 5mm 90mm, clip]{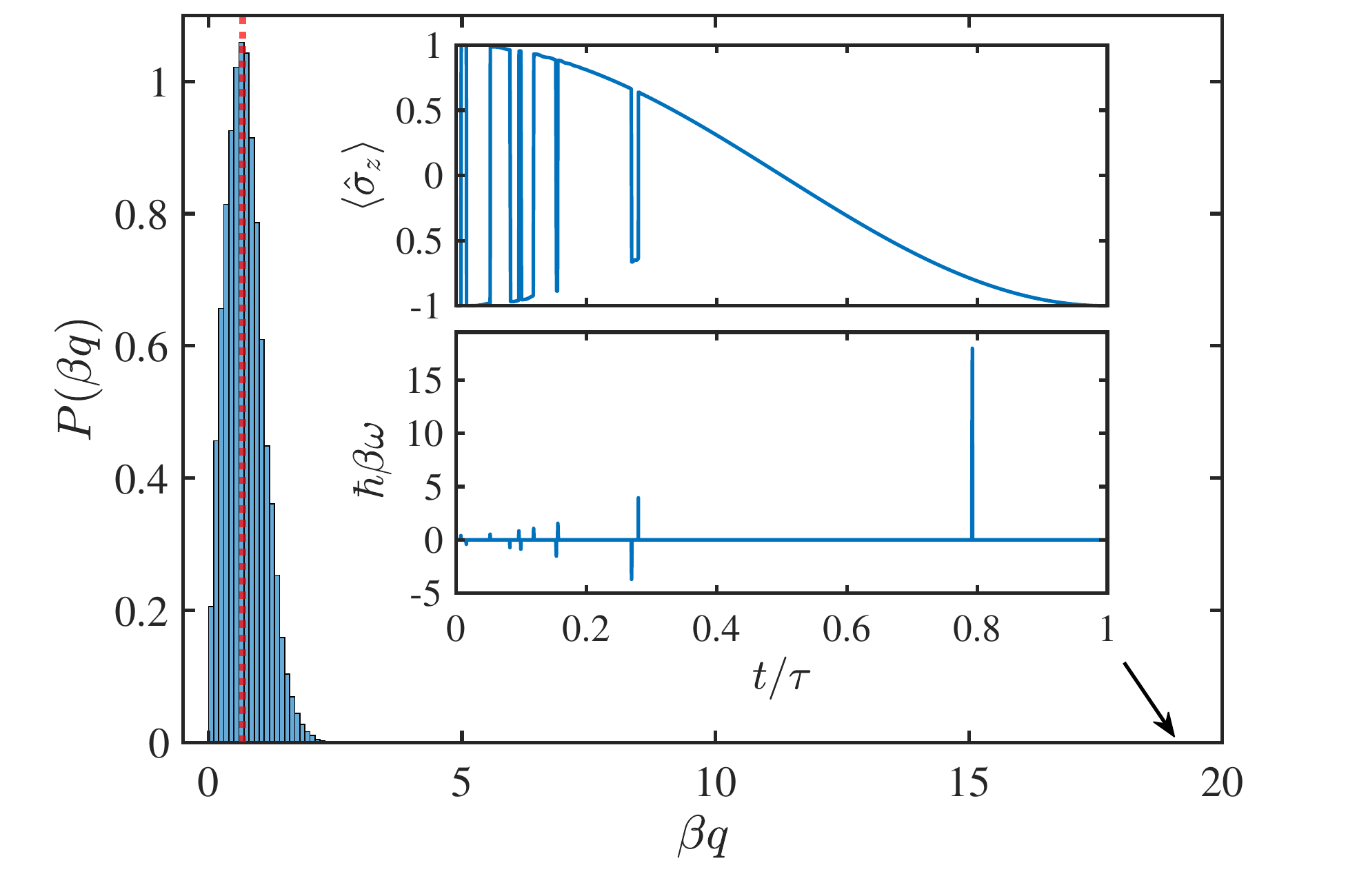}
    \caption{Quantum-jump trajectory simulation of the coherent qubit erasure protocol of Fig.~\ref{fig:cumulants}, with $\beta\varepsilon_\tau = 20$. Main panel: heat distribution over $3\times 10^4$ trajectories, with the Landauer bound $\beta q = -\Delta S$ shown by the red dotted line. Inset: dynamics of a rare trajectory with large heat transfer, $q =19.1 k_B T$ (black arrow). Stochastic jumps in the otherwise continuous evolution of $\langle \hat{\sigma}_z\rangle$ (upper inset) are associated with the emission of energy quanta $\hbar\omega = \pm \varepsilon_t$ to the environment (lower inset). Non-adiabatic quantum evolution allows for two consecutive emissions and consequently extreme dissipation.\label{fig:distribution_outliers}}
\end{figure}

To reveal the microscopic origin of these tails, we simulate individual runs of the erasure protocol using the quantum-jump trajectory approach (see Appendix~\ref{app:trajectories}). A trajectory is described by a pure state $\ket{\psi_t}$ undergoing continuous time evolution interspersed by stochastic jumps, $\ket{\psi_t}\to \ket{\pm\varepsilon_t}$, where $\hat{H}_t\ket{\pm\varepsilon_t} = \pm\tfrac{1}{2}\varepsilon_t\ket{\pm \varepsilon_t}$. Each jump transfers a quantum of energy $\hbar\omega = \mp\varepsilon_t$ to the environment. The main panel of Fig.~\ref{fig:distribution_outliers} shows the heat distribution obtained by numerically sampling many such trajectories for a quantum process. While the bulk of the distribution is centred around the Landauer bound, we find a few rare trajectories featuring a very large heat transfer, which are associated with non-adiabatic transitions occurring during the driving protocol. For example, consider an emission at some time, which leaves the system in its instantaneous ground state. As the eigenbasis of the Hamiltonian rotates, the state at some later time comprises a superposition of both energy eigenstates. The finite population of the excited state thus opens the possibility for a second emission to occur, potentially leading to massive overall heat transfer. An example of such a trajectory is shown in the inset of Fig.~\ref{fig:distribution_outliers}. On the contrary, during a classical protocol the state adiabatically follows the Hamiltonian eigenstates between jumps. This implies that an emission can only be followed by an absorption and vice versa, such that the contributions of these alternating events to the heat statistics largely cancel. We note that, apart from these rare trajectories, the heat distributions sampled from quantum protocols are very similar to their classical counterparts, with the bulk of the distribution approaching a Gaussian form as $\tau$ increases (see Appendix~\ref{app:trajectories}, Fig.~\ref{fig:quantum_classical_distributions}). The excess skewness and kurtosis of the quantum heat distributions can therefore be attributed entirely to rare, non-adiabatic processes such as the one illustrated in Fig.~\ref{fig:distribution_outliers}.

Even though such events are statistical outliers, they may have severe consequences for nanoscale heat management. For the data shown in Fig.~\ref{fig:distribution_outliers}, roughly one trajectory in every thousand involves a non-adiabatic transition. However, the maximum heat dissipated in a single trajectory is more than 30 times larger than the Landauer limit, whereas for the analogous classical protocol it is less than four times larger. This illustrates that quantum coherence drastically increases the probability of extreme heat fluctuations during the process of information erasure. Such events could damage or disrupt small-scale quantum hardware with a low threshold of tolerance for heat dissipation. These are truly quantum fluctuations, in the sense that uncertainty in the transferred heat is increased by the existence of a coherent superposition state of the system together with the quantisation of energy exchanged with the environment. In the context of qubit erasure, these quantum fluctuations are experimentally distinguishable from thermal fluctuations since only the former involve consecutive emission or absorption events.

The results presented here can be applied to other logic operations implemented on physical hardware. Indeed, we expect that unique energetic fingerprints may also be discovered in other control protocols that process information in the quantum regime. Fast protocols that push the system far from equilibrium~\cite{Proesmans2020} are especially important for computing at high clock speed but are also expected to incur even greater heat fluctuations. Recently developed methods to describe dissipation in driven open quantum systems~\cite{Dann2019,Dann2020,Pancotti2020,Popovic2020} could be used to address this problem in future work.

\textit{Acknowledgments.---} We thank Sebastian Deffner, G\'eraldine Haack, Nicole Yunger Halpern, Kavan Modi and Mart\'i Perarnau-Llobet for useful comments on the manuscript. We acknowledge support from the European Research Council Starting Grant ODYSSEY (G. A. 758403). Calculations were performed on the Lonsdale cluster maintained by the Trinity Centre for High Performance Computing. This cluster was funded through grants from Science Foundation Ireland. JG acknowledges support from a SFI-Royal Society University Research Fellowship.

\bibliographystyle{apsrev4-1}
\bibliography{mybib2}

\newpage

\widetext

\appendix

\setcounter{secnumdepth}{2}
\setcounter{figure}{0}
\renewcommand{\thefigure}{\thesection\arabic{figure}}

\section{Proof of~\eqref{eq:split}}
\label{app:split}
\

\noindent 

In this Appendix, we will show how to decompose the CGF for excess heat into a quantum and classical contribution. As stated in the main text, we assume that the quantum Markovian semigroup governing the open system dynamics satisfies the detailed balance condition  $\mathscr{L}_t^*(.)=\tilde{\mathscr{L}}_t(.)+2i [\hat{H}_t,(.)]$, where $\tilde{\mathscr{L}}_t$ is a dual generator determined by $\tr{\tilde{\mathscr{L}}_t(\hat{A})\hat{\pi}_t  \hat{B}}=\tr{\hat{A} \hat{\pi}_t \mathscr{L}_t^*(\hat{B})}$ for any bounded operators $\hat{A},\hat{B}$ \cite{Alicki1976}. Here, and throughout this Supplemental Material, we set $\hbar=1$. We also assume $\hat{H}_t$ is non-degenerate for simplicity. Note that while this condition remains valid at the beginning of the protocol, the difference in energy eigenvalues are negligible when compared to the temperature of the environment. It has been proven that this condition ensures that $\tilde{\mathscr{L}}$ also forms a quantum Markovian semigroup (Lemma 5.1, Ref.~\cite{Fagnola2007}), which in turn implies that the generator satisfies the following time-translational symmetry (Theorem 3.1, Ref.~\cite{Fagnola2007}):
\begin{align}
    \mathscr{L}^*_t(.)=\hat{\pi}_t^{i\alpha}\mathscr{L}^*_t \bigg(\hat{\pi}_t^{-i\alpha}(.)\hat{\pi}_t^{i\alpha}\bigg)\hat{\pi}_t^{-i\alpha}, \ \ \ \ \forall \alpha\in\mathbb{R}.
\end{align}
Since the fixed point is thermal, and hence a function of the Hamiltonian, we may write this condition in the equivalent form
\begin{align}\label{eq:time_cov}
    [\mathscr{L}^*_t,\mathscr{H}_t]=0,
\end{align}
where $\mathscr{H}_t(.)=[\hat{H}_t,(.)]$. A useful identity for the dephasing operation $\mathcal{D}_{\hat{H}_t}(.)=\sum_{n}\ket{n_t}\bra{n_t}(.) \ket{n_t}\bra{n_t}$ is via the infinite time-average for non-degenerate Hamiltonians, namely
\begin{align}\label{eq:time_av}
    \mathcal{D}_{\hat{H}_t}(.)=\lim_{\mathcal{T}\to\infty}\frac{1}{\mathcal{T} }\int^{\mathcal{T}}_0 d\mu \ e^{-i\mu \mathscr{H}_t}(.),
\end{align}
Using~\eqref{eq:time_cov} this implies commutation between the adjoint Lindblad evolution (at fixed $t$) and the dephasing operation, i.e.
\begin{align}\label{eq:dephase}
    [\mathcal{D}_{\hat{H}_t},e^{\nu\mathscr{L}^*_t}]=[\mathcal{D}_{\hat{H}_t},e^{\nu\mathscr{L}_t}]=0, 
\end{align}

To quantify the slow-driving regime, from here on we rescale the variable for time so that $ \epsilon t\to t\in[0,1]$ with $\epsilon=1/\tau$ denoting the driving speed given by the inverse of the total time of the protocol. In these coordinates we can express the system state in a form $\rho_{t}=\pi_{t}+\epsilon\delta \rho_{t} +\mathcal{O}(\epsilon^2)$. Under the slow-driving approximation one may derive a compact expression for the CGF [Eq.~\eqref{eq:CGF}] to leading order in driving speed (see Ref.~\cite{Scandi2020}):
\begin{align}\label{eq:CGF2}
    \mathcal{K}_q(u)\simeq-\epsilon \ \beta^2\int^1_0 dt  \int^\infty_0 d\nu \ \big[\hspace{-2mm}\big[\,  \text{cov}^u_{t}\big(\dot{\hat{H}}_{t}(\nu),\dot{\hat{H}}_{t}\big)\big] \hspace{-2mm}\big]\,,
\end{align}
where we denote Heisenberg-evolved operators by $\hat{A}_{t}(\nu)=e^{\nu \mathscr{L}^{*}_{t}}(\hat{A}_{t})$ at a frozen value $t$, and
\begin{align}\label{eq:y_cov}
\text{cov}^y_t(\hat{A},\hat{B})=\tr{\hat{A}\hat{\pi}_t^y \hat{B} \hat{\pi}_t^{1-y}}-\tr{\hat{A}\hat{\pi}_t}\tr{\hat{B}\hat{\pi}_t},
\end{align}
is known as the quantum covariance, a non-commutative generalisation of the classical covariance $\langle a b\rangle-\langle a \rangle\langle b\rangle$. Finally, in Eq.~\eqref{eq:CGF2} we denote $[\![  f(u) ]\!]=\int^{u k_B T}_0 dy \int_y^{1-y} dy' \ f(y')$ as a particular averaging over the counting field $u$ at temperature $T$. Note that our approximation is valid as long as the protocol duration $\tau$ is much larger than the relaxation time associated with $\mathscr{L}_t$ at each time. The proof of~\eqref{eq:CGF2} can be obtained via two equivalent routes. Firstly, it was derived in Ref.~\cite{Scandi2020} by applying a weak coupling assumption to the global unitary evolution of system and bath, and expanding the resulting entropy production CGF in terms of the Hamiltonian driving speed. For Landauer erasure considered here, the statistics of entropy production become equivalent to the dissipated heat statistics. Alternatively, another derivation was provided in \cite{Miller2020a} by directly applying adiabatic perturbation theory to the CGF obtained via a quantum jump approach. 

To see how coherence impacts the heat statistics, we partition the operator $\dot{\hat{H}}_t$ into a classical (diagonal) and quantum-coherent part respectively, $\dot{\hat{H}}_t=\dot{\hat{H}}_t^d+\dot{\hat{H}}^c_t$, with
\begin{align}\label{eq:decomp}
    \dot{\hat{H}}^d_t=\mathcal{D}_{\hat{H}_t}(\dot{\hat{H}}_t), \ \ \ \ \ \dot{\hat{H}}^c_t=\dot{\hat{H}}_t-\dot{\hat{H}}^d_t,
\end{align}
where $\mathcal{D}_{\hat{H}_t}(.)=\sum_{n}\ket{n_t}\bra{n_t}(.) \ket{n_t}\bra{n_t}$ denotes the dephasing map in the instantaneous energy eigenbasis $\{ \ket{n_t}\}$ of $\hat{H}_t$. The operator $\dot{\hat{H}}^c_t$ represents the effect of quantum non-commutativity as it vanishes only if the Hamiltonian commutes with itself at all different times throughout the protocol. Such an approach has previously been used to identify non-classical signatures in quantum heat engines \cite{Brandner2017} and quenched quantum systems \cite{Scandi2020}.

We now expand the quantum covariance in~\eqref{eq:CGF2} and substitute in the decomposition of the power operator in Eq.~\eqref{eq:decomp}:
\begin{align}\label{eq:covexpand}
     \nonumber\text{cov}^u_t\big(\dot{\hat{H}}_t(\nu),\dot{\hat{H}}_t\big)&=\text{cov}^u_t\big(e^{\nu\mathscr{L}^*_t}\big(\dot{\hat{H}}_t),\dot{\hat{H}}_t\big) \\ 
     \nonumber&=\text{cov}^u_t\big(e^{\nu\mathscr{L}^*_t}\big(\dot{\hat{H}}_t^d+\dot{\hat{H}}^c_t),\dot{\hat{H}}_t^d+\dot{\hat{H}}^c_t\big) \\
     &=\text{cov}^u_t\big(\dot{\hat{H}}_t^d(\nu),\dot{\hat{H}}_t^d\big)+\text{cov}^u_t\big(\dot{\hat{H}}^c_t(\nu),\dot{\hat{H}}^c_t\big)+\text{cov}^u_t\big(\dot{\hat{H}}^c_t(\nu),\dot{\hat{H}}^d_t\big)+\text{cov}^u_t\big(\dot{\hat{H}}^d_t(\nu),\dot{\hat{H}}^c_t\big),
\end{align}
where we used the linearity of the map $e^{\nu\mathscr{L}^*_t}$. It is straightforward to show that the coherent power has zero diagonal elements so that $\bra{n_t}\dot{\hat{H}}^c_t\ket{n_t}=0$ for any energy eigenstate $n_t$. This means that
\begin{align}\label{eq:dephasezero}
    \mathcal{D}_{\hat{H}_t}(\dot{\hat{H}}^c_t)=0.
\end{align}
Looking at the cross terms in~\eqref{eq:covexpand}, one sees that they vanish:
\begin{align}
    \nonumber\text{cov}^u_t\big(\dot{\hat{H}}^c_t(\nu),\dot{\hat{H}}^d_t\big)&=\tr{e^{\nu \mathscr{L}_t^*}(\dot{\hat{H}}^c_t)\hat{\pi}_t^u \mathcal{D}_{\hat{H}_t}(\dot{\hat{H}}_t-\langle \dot{\hat{H}}_t \rangle)\hat{\pi}^{1-u}} \\
    \nonumber&=\tr{e^{\nu \mathscr{L}_t^*}(\dot{\hat{H}}^c_t) \mathcal{D}_{\hat{H}_t}\big(\hat{\pi}_t^u(\dot{\hat{H}}_t-\langle \dot{\hat{H}}_t \rangle)\hat{\pi}^{1-u}\big)} \\
    \nonumber&=\tr{\mathcal{D}_{\hat{H}_t}\big(e^{\nu \mathscr{L}_t^*}(\dot{\hat{H}}^c_t)\big) \hat{\pi}_t^u(\dot{\hat{H}}_t-\langle \dot{\hat{H}}_t \rangle)\hat{\pi}^{1-u}} \\
    \nonumber&=\text{cov}^u_t\big(\mathcal{D}_{\hat{H}_t}\big(e^{\nu \mathscr{L}_t^*}(\dot{\hat{H}}^c_t)\big),\dot{\hat{H}}_t\big), \\
    \nonumber&=\text{cov}^u_t\big(e^{\nu \mathscr{L}_t^*}\mathcal{D}_{\hat{H}_t}(\dot{\hat{H}}^c_t),\dot{\hat{H}}_t\big) \\
    &=0,
\end{align}
where in the first line we have rewritten Eq.~\eqref{eq:y_cov} using the thermal average $\langle \dot{\hat{H}}_t\rangle = \Tr{\dot{\hat{H}}_t\hat{\pi}_t}$, and we used~\eqref{eq:dephase} in the fourth line and~\eqref{eq:dephasezero} in the fifth line. Following the same steps we can also show 
\begin{align}\label{eq:crossterm}
    \text{cov}^u_t\big(\dot{\hat{H}}^d_t(\nu),\dot{\hat{H}}^c_t\big)=0,
\end{align}
With the vanishing cross terms~\eqref{eq:covexpand} we find that the quantum covariance divides into a classical and quantum part respectively:
\begin{align}
    \text{cov}^u_t\big(\dot{\hat{H}}_t(\nu),\dot{\hat{H}}_t\big)=\text{cov}^u_t\big(\dot{\hat{H}}_t^d(\nu),\dot{\hat{H}}_t^d\big)+\text{cov}^u_t\big(\dot{\hat{H}}_t^c(\nu),\dot{\hat{H}}_t^c\big),
\end{align}
Substituting this into Eq.~\eqref{eq:CGF2} completes the proof of Eq.~\eqref{eq:split}, with 
\begin{align}\label{eq:class}
    \mathcal{K}^d_q(u) & =\epsilon ( u^2-\beta u)\int^1_0 dt \int^\infty_0 d\nu \ \text{cov}_t\big(\dot{\hat{H}}^d_{t}(\nu),\dot{\hat{H}}^d_{t}\big), \\
\label{eq:quant}
    \mathcal{K}^c_q(u) &=-\epsilon \beta^2\int^1_0 dt \ \int^\infty_0 d\nu \ \big[\hspace{-2mm}\big[\, \text{cov}^u_t\big(\dot{\hat{H}}^c_{t}(\nu),\dot{\hat{H}}^c_{t}\big)\big]\hspace{-2mm}\big].
\end{align}
Here we have introduced the symmeterised covariance 
\begin{align}
    \text{cov}_t\big(\hat{A},\hat{B}\big):=\frac{1}{2}\tr{\{\hat{A},\hat{B}\}\hat{\pi}_t  }-\tr{\hat{A}\hat{\pi}_t}\tr{\hat{B}\hat{\pi}_t}
\end{align}
One sees that the CGF is composed of two correlation functions; a classical contribution $\mathcal{K}^d_q(u)$ given by the correlations in the diagonal operator $\dot{\hat{H}}_t^d$, alongside a quantum contribution $\mathcal{K}^c_q(u)$ given by the correlations in the non-diagonal operator $\dot{\hat{H}}_t^d$. As we will later see in the next section, these functions can be related directly to the change in diagonal and off-diagonal elements of the system density operator in the energy basis.  

\section{Heat cumulants and Renyi divergences}
\label{app:Renyi}
\

In this appendix we will demonstrate that the two CGF's~\eqref{eq:class} and ~\eqref{eq:quant} can be equivalently expressed in terms of the quantum Renyi divergences~\eqref{eq:CGFd_renyi} and~\eqref{eq:CGFq_renyi}. Firstly, when expressed in in terms of the rescaled time $\epsilon t\to t\in[0,1]$, the dynamics of the system obey the master equation
\begin{align}\label{eq:master2}
    \dot{\hat{\rho}}_t=\epsilon^{-1}\mathscr{L}_t(\hat{\rho}_t).
\end{align}
For slow driving we assume $\tau=\epsilon^{-1}\gg \tau^{\rm eq}$, where $\tau^{\rm eq}$ is the characteristic timescale determined by the eigenvalues of the generator $\mathscr{L}_t$. We then expect the solution to~\eqref{eq:master2} to remain close to the fixed point $\hat{\pi}_t$. To demonstrate this we will need to utilise the following Taylor expansion of the system density matrix up to first order in driving speed  \cite{Cavina2017,Scandi,Harry}:
\begin{align}\label{eq:delta_rho}
    \hat{\rho}_t=\hat{\pi}_t+\epsilon \delta \hat{\rho}_t+\mathcal{O}(\epsilon^2),
\end{align}
where
\begin{align}
    \delta \hat{\rho}_t:=-\beta \mathcal{L}^+_t \mathbb{J}_{\hat{\pi}_t}(\Delta \dot{\hat{H}}_t).
\end{align}
with $\Delta \dot{\hat{H}}_t=\dot{\hat{H}}_t-\tr{\hat{\pi}_t \dot{\hat{H}}_t}$. Here we have introduced the \textit{Drazin inverse} of the Lindbladian, defined by
\begin{align}\label{eq:drazin}
    \mathscr{L}^{+}_t(.):=-\int^\infty_0 d\nu \ e^{\nu \mathscr{L}_t}[(.)-\hat{\pi}_t\tr{.}],
    \end{align}
Note here that the Drazin inverse acts as
\begin{align}\label{eq:inverse_draz}
    \mathscr{L}_t^+\mathscr{L}_t(\hat{A})=\mathscr{L}_t\mathscr{L}^+_t(\hat{A})=\hat{A}-\tr{\hat{A}}\hat{\pi}_t.
\end{align}
Furthermore we have introduced the logarithmic matrix mean \cite{Petz2014}
\begin{align}\label{eq:log_mean}
    \mathbb{J}_{\hat{\rho}} (.):=\int^1_0 dx \ \hat{\rho}^x (.) \hat{\rho}^{1-x}.
\end{align}
Note that for the dephased density matrix $\bar{\hat{\rho}}_t=\mathcal{D}_{\hat{H}_t}(\hat{\rho}_t)$ we also have
\begin{align}\label{eq:delta_bar}
    \bar{\hat{\rho}}_t=\hat{\pi}_t+\epsilon \delta \bar{\hat{\rho}}_t+\mathcal{O}(\epsilon^2), \ \ \ \text{with} \ \ \ \delta\bar{\hat{\rho}}_t=-\beta \mathcal{L}^+_t \mathbb{J}_{\hat{\pi}_t}(\Delta \dot{\hat{H}}_t^d ).
\end{align}
Here we have used the fact that 
\begin{align}
    [\mathscr{L}^+_t, \mathcal{D}_{\hat{H}_t}]=[\mathbb{J}_{\hat{\pi}_t}, \mathcal{D}_{\hat{H}_t}]=0,
\end{align}
The first commutation relation can be verified by combining~\eqref{eq:drazin} with the condition of time-translational covariance~\eqref{eq:dephase} together with the fact that the dephasing map is trace-preserving. The second commutation relation can be verified by using the representation~\eqref{eq:time_av} along with $[\hat{\pi}_t,\hat{H}_t]=0$. Before proceeding it will be useful to use the following Taylor expansion of the matrix logarithm \cite{Petz2014}: 
\begin{align}\label{eq:log_exp}
    \ln{\hat{\rho}+\epsilon \hat{\hat{\sigma}}}=\ln{\hat{\rho}}+\epsilon \mathbb{J}_{\hat{\rho}}^{-1}(\hat{\sigma})+\mathcal{O}(\epsilon^2),
\end{align}
where
\begin{align}\label{eq:log_exp_int}
    \mathbb{J}_{\hat{\rho}}^{-1}(.):=\int^\infty_0 dx \ (\hat{\rho}+x \id)^{-1} (.)(\hat{\rho}+x\id)^{-1}.
\end{align}
It is important to note that both $\mathbb{J}_{\hat{\rho}}(.)$ and $\mathbb{J}_{\hat{\rho}}^{-1}(.)$ are hermitian with respect to the Hilbert-Schmidt inner product $\langle \hat{A},\hat{B} \rangle=\tr{\hat{B}^\dagger \hat{A}}$ and are inverse to each other, i.e. 
\begin{align}
    \mathbb{J}_{\hat{\rho}} \ \mathbb{J}^{-1}_{\hat{\rho}}(\hat{A})=\mathbb{J}^{-1}_{\hat{\rho}} \mathbb{J}_{\hat{\rho}}(\hat{A})=\hat{A}.
\end{align}
We will begin by expanding the derivative of the relative entropy between the dephased state and the fixed point, namely 
\begin{align}\label{eq:dot_q_diag}
    \frac{\partial}{\partial t}S(\mathcal{D}_{\hat{H}_s}(\hat{\rho}_t) || \hat{\pi}_{s})\bigg|_{s=t}=-\epsilon^{-1} \ \tr{\mathscr{L}_t(\bar{\hat{\rho}}_t)\big(\ln{\bar{\hat{\rho}}_t}-\ln{\hat{\pi}_t}\big)}.
\end{align}
Note that the second equality follows again from the commutation $[\mathscr{L}_t,\mathcal{D}_{\hat{H}_t}]=0$. We next substitute in~\eqref{eq:delta_bar} and keep term only up to second order in $\epsilon$:
\begin{align}
    \nonumber\tr{\mathscr{L}_t(\bar{\hat{\rho}}_t)\big(\ln{\bar{\hat{\rho}}_t}-\ln{\hat{\pi}_t}\big)}&\simeq\epsilon^2\beta^2\tr{\mathscr{L}_t \mathscr{L}^+_t \mathbb{J}_{\hat{\pi}_t}(\Delta \dot{\hat{H}}^d_t )\mathbb{J}^{-1}_{\hat{\pi}_t}\mathscr{L}_t^{+}\mathbb{J}_{\hat{\pi}_t}(\Delta \dot{\hat{H}}_t^d) }, \\
    \nonumber&=\epsilon^2\beta^2\tr{\mathbb{J}^{-1}_{\hat{\pi}_t}\mathbb{J}_{\hat{\pi}_t}(\Delta \dot{\hat{H}}^d_t )\mathscr{L}_t^{+}\mathbb{J}_{\hat{\pi}_t}(\Delta \dot{\hat{H}}_t^d) }, \\
    \nonumber&=\epsilon^2\beta^2\tr{\Delta \dot{\hat{H}}^d_t \mathscr{L}_t^{+}\mathbb{J}_{\hat{\pi}_t}(\Delta \dot{\hat{H}}_t^d) }, \\
    \nonumber&=-\epsilon^2\beta^2\int^\infty_0 d\nu \ \tr{e^{\nu \mathscr{L}^*_t}(\dot{\hat{H}}^d_t) \mathbb{J}_{\hat{\pi}_t}(\Delta \dot{\hat{H}}_t^d) }, \\ 
    \nonumber&=-\epsilon^2\beta^2 \int^\infty_0 d\nu \int^1_0 dx \ \text{cov}^x_t(\dot{\hat{H}}_t^d(\nu),\dot{\hat{H}}_t^d), \\
    &=-\epsilon^2\beta^2 \int^\infty_0 d\nu  \ \text{cov}_t(\dot{\hat{H}}_t^d(\nu),\dot{\hat{H}}_t^d),
\end{align}
where in the second line we used~\eqref{eq:inverse_draz} and the fact that $\mathbb{J}^{-1}_{\hat{\rho}}$ is hermitian, in the fourth line we used the definition~\eqref{eq:drazin}, and in the final line we used $[\dot{\hat{H}}_t^d,\hat{\pi}_t]=0$. We thus conclude
\begin{align}
    \frac{\partial}{\partial t}S(\mathcal{D}_{\hat{H}_s}(\hat{\rho}_t) || \hat{\pi}_{s})\bigg|_{s=t}=-\epsilon\beta^2 \int^\infty_0 d\nu  \ \text{cov}_t(\dot{\hat{H}}_t^d(\nu),\dot{\hat{H}}_t^d),
\end{align}
which proves~\eqref{eq:CGFd_renyi} in the main text. We next focus on expanding the time derivative of the quantum Renyi divergence between the system state and its dephased counterpart:
\begin{align}
    (\alpha -1) \dot{S}_\alpha(\hat{\rho}_t || \bar{\hat{\rho}}_t(s))=\frac{d}{d t}\text{ln} \ \tr{\hat{\rho}_t^\alpha \bar{\hat{\rho}}^{1-\alpha}_t(s)}, \ \ \ \ \alpha\in(0,1)\cup(1,\infty),
\end{align}
where we set $\bar{\hat{\rho}}_t(s)=\mathcal{D}_{\hat{H}_s}(\hat{\rho}_t)$ and denote $\bar{\hat{\rho}}_t(s=t)=\bar{\hat{\rho}}_t$. We will adapt a method presented in \cite{Scandi2020} that was used to expand Renyi divergences by first expanding the trace as
\begin{align}
    \tr{\hat{\rho}_t^\alpha \bar{\hat{\rho}}^{1-\alpha}_t(s)}=1+\int^\alpha_0 dx \ \tr{\hat{\rho}_t^x\big(\text{ln} \ \hat{\rho}_t-\text{ln} \ \bar{\hat{\rho}}_t(s)\big)\bar{\hat{\rho}}^{1-x}_t(s)}.
\end{align}
We then have
\begin{align}\label{eq:renyi1}
   \nonumber\frac{d}{dt}\text{ln} \ \tr{\hat{\rho}_t^\alpha \bar{\hat{\rho}}^{1-\alpha}_t(s)}&=\lim_{\Delta t\to 0}\Delta t^{-1} \tr{\hat{\rho}_t^\alpha \bar{\hat{\rho}}^{1-\alpha}_t(s)}^{-1}\text{ln} \ \tr{\hat{\rho}_{t+\Delta t}^\alpha \bar{\hat{\rho}}^{1-\alpha}_{t+\Delta t}(s)} , \\
   \nonumber&=\lim_{\Delta t\to 0}\Delta t^{-1} \tr{\hat{\rho}_t^\alpha \bar{\hat{\rho}}^{1-\alpha}_t(s)}^{-1}\text{ln} \ \bigg(1+\Delta t\int^\alpha_0 dx \ \frac{d}{d t} \tr{\hat{\rho}_t^x\big(\text{ln} \ \hat{\rho}_t-\text{ln} \ \bar{\hat{\rho}}_t(s)\big)\bar{\hat{\rho}}^{1-x}_t(s)}+\mathcal{O}(\Delta t^2)\bigg), \\
   &=\tr{\hat{\rho}_t^\alpha \bar{\hat{\rho}}^{1-\alpha}_t(s)}^{-1}\int^\alpha_0 dx \ \frac{d}{d t} \tr{\hat{\rho}_t^x\big(\text{ln} \ \hat{\rho}_t-\text{ln} \ \bar{\hat{\rho}}_t(s)\big)\bar{\hat{\rho}}^{1-x}_t(s)},
\end{align}
where we used $\ln{1+x}=x+\mathcal{O}(x^2)$ to evaluate the limit. To proceed we will require the following expansion \cite{Scandi2020}:
\begin{align}\label{eq:log_x_exp}
    (\hat{\rho}+\epsilon \hat{\sigma})^x=\hat{\rho}^x+\epsilon\int^x_0 dy \ \hat{\rho}^y \mathbb{J}^{-1}_{\hat{\rho}}( \hat{\sigma})\hat{\rho}^{x-y}+\mathcal{O}(\epsilon^2),
\end{align}
where $\mathbb{J}^{-1}_{\hat{\rho}}$ is defined in~\eqref{eq:log_exp_int}. We now introduce the slow driving expansions $\hat{\rho}_t=\hat{\pi}_t+\epsilon \delta \hat{\rho}_t+\mathcal{O}(\epsilon^2)$ and $\bar{\hat{\rho}}_t=\hat{\pi}_t+\epsilon \delta \bar{\hat{\rho}}_t+\mathcal{O}(\epsilon^2)$ defined in~\eqref{eq:delta_rho} and~\eqref{eq:delta_bar} respectively. Then from~\eqref{eq:log_x_exp} we find
\begin{align}\label{eq:prefactor}
    \nonumber\tr{\hat{\rho}_t^\alpha \bar{\hat{\rho}}_t^{1-\alpha}}&=\tr{\hat{\pi}_t^\alpha \bar{\hat{\rho}}_t^{1-\alpha}}+\epsilon\int^\alpha_0 dy \ \tr{\hat{\pi}_t^y \mathbb{J}^{-1}_{\hat{\pi}_t}(\delta\hat{\rho}_t)\hat{\pi}_t^{\alpha-y}\bar{\hat{\rho}}_t^{1-\alpha}}+\mathcal{O}(\epsilon^2), \\
    \nonumber&=1+\epsilon\int^{1-\alpha}_0 dy \ \tr{\hat{\pi}_t^{\alpha+y} \mathbb{J}^{-1}_{\hat{\pi}_t}(\delta\bar{\hat{\rho}}_t)\hat{\pi}_t^{1-\alpha-y}}+\epsilon\int^\alpha_0 dy \ \tr{\hat{\pi}_t^{1-y} \mathbb{J}^{-1}_{\hat{\pi}_t}(\delta\hat{\rho}_t)\hat{\pi}_t^{1-y}}+\mathcal{O}(\epsilon^2), \\
    \nonumber&=1+\epsilon(1-\alpha)\tr{\hat{\pi}_t\mathbb{J}^{-1}_{\hat{\pi}_t}(\delta\bar{\hat{\rho}}_t)}+\epsilon\alpha\tr{\hat{\pi}_t\mathbb{J}^{-1}_{\hat{\pi}_t}(\delta\hat{\rho}_t)}+\mathcal{O}(\epsilon^2), \\
    \nonumber&=1+\epsilon\tr{\mathbb{J}^{-1}_{\hat{\pi}_t}(\hat{\pi}_t)\delta\hat{\rho}_t}+\epsilon\tr{\mathbb{J}^{-1}_{\hat{\pi}_t}(\hat{\pi}_t)\delta\bar{\hat{\rho}}_t}+\mathcal{O}(\epsilon^2), \\
    &=1+\mathcal{O}(\epsilon^2),
\end{align}
where in the third line we used the cyclicity of the trace, in the fourth line we used the hermicity of $\mathbb{J}^{-1}_{\hat{\pi}_t}$, and in the final line we used $\mathbb{J}^{-1}_{\hat{\pi}_t}(\hat{\pi}_t)=\id$ together with the fact that $\tr{\delta \hat{\rho}_t}=\tr{\delta \bar{\hat{\rho}}_t}=0$. If we compare this with~\eqref{eq:renyi1}, we see that the integrand vanishes at zero order in $\epsilon$ at $t=s$. Therefore under slow driving we can neglect the prefactor~\eqref{eq:prefactor} in front, giving
\begin{align}\label{eq:renyi2}
    \nonumber&\frac{d}{dt}\text{ln} \ \tr{\hat{\rho}_t^\alpha \bar{\hat{\rho}}^{1-\alpha}_t(s)}\bigg|_{s=t}\simeq\int^\alpha_0 dx \ \frac{d}{dt} \tr{\hat{\rho}_t^x\big(\text{ln} \ \hat{\rho}_t-\text{ln} \ \bar{\hat{\rho}}_t(s)\big)\bar{\hat{\rho}}^{1-x}_t(s)}\bigg|_{s=t}, \\
    &=\int^\alpha_0 dx \ A(x)+B(x)+C(x),
\end{align}
where for convenience we have defined 
\begin{align}
    &A(x):=\tr{\bigg(\frac{d}{dt} \hat{\rho}_t^x\bigg)\big(\text{ln} \ \hat{\rho}_t-\text{ln} \ \bar{\hat{\rho}}_t(s)\big)\bar{\hat{\rho}}^{1-x}_t(s)}\bigg|_{s=t}, \\
    &B(x):=\tr{ \hat{\rho}_t^x\frac{d}{dt}\bigg(\text{ln} \ \hat{\rho}_t-\text{ln} \ \bar{\hat{\rho}}_t(s)\bigg)\bar{\hat{\rho}}^{1-x}_t(s)}\bigg|_{s=t}, \\
    &C(x):=\tr{ \hat{\rho}_t^x\big(\text{ln} \ \hat{\rho}_t-\text{ln} \ \bar{\hat{\rho}}_t(s)\big)\bigg(\frac{d}{dt}\bar{\hat{\rho}}^{1-x}_t(s)\bigg)}\bigg|_{s=t}, 
\end{align}
We now expand each term up to first order in $\epsilon$, in which case we first find
\begin{align}
    \nonumber A(x)&=\epsilon^{-1}\int^x_0 dy \ \tr{\hat{\rho}_t^y \mathbb{J}_{\hat{\rho}_t}^{-1}\mathscr{L}_t (\hat{\rho}_t)\hat{\rho}_t^{x-y}\big(\text{ln} \ \hat{\rho}_t-\text{ln} \ \bar{\hat{\rho}}_t\big)\bar{\hat{\rho}}^{1-x}_t}, \\
    \nonumber&=-\beta\epsilon\int^x_0 dy \ \tr{\Delta \dot{\hat{H}}_t \hat{\pi}_t^{x-y}\big(\mathbb{J}_{\hat{\pi}_t}^{-1}(\delta\hat{\rho}_t-\delta \bar{\hat{\rho}}_t)\big)\hat{\pi}^{1+y-x}_t}+\mathcal{O}(\epsilon^2), \\
    \nonumber&=\epsilon\beta^2\int^x_0 dy \ \tr{\Delta \dot{\hat{H}}_t \hat{\pi}_t^{x-y}\mathbb{J}_{\hat{\pi}_t}^{-1}\mathscr{L}^+_t \mathbb{J}_{\hat{\pi}_t}(\dot{\hat{H}}^c_t) \hat{\pi}^{1+y-x}_t}+\mathcal{O}(\epsilon^2), \\
    &=\epsilon\beta^2\int^x_0 dy \ \tr{\mathbb{J}_{\hat{\pi}_t}^{-1}\big(\hat{\pi}^{1-(x-y)}_t\Delta \dot{\hat{H}}_t \hat{\pi}_t^{x-y}\big)\mathscr{L}^+_t \mathbb{J}_{\hat{\pi}_t}(\dot{\hat{H}}^c_t) }+\mathcal{O}(\epsilon^2), 
\end{align}
where we used~\eqref{eq:log_x_exp} in the first line,~\eqref{eq:log_exp} together with~\eqref{eq:log_exp_int} and~\eqref{eq:inverse_draz} in the second line, and the hermicity of $\mathbb{J}^{-1}_{\hat{\rho}}$ in the final line. Integration over $x$ and performing a change of variables $x'=x$ and $y'=x-y$ yields
\begin{align}\label{eq:a_part}
    \nonumber\int^\alpha_0 dx \ A(x)&=-\epsilon\beta^2\int^\alpha_0 dx' \int^{x'}_0 dy' \ \tr{\mathbb{J}_{\hat{\pi}_t}^{-1}\big(\hat{\pi}^{1-y'}_t\Delta \dot{\hat{H}}_t \hat{\pi}_t^{y'}\big)\mathscr{L}^+_t \mathbb{J}_{\hat{\pi}_t}(\dot{\hat{H}}^c_t) }+\mathcal{O}(\epsilon^2), \\
    \nonumber&=-\epsilon\beta^2\int^\alpha_0 dx' \int^{x'}_0 dy' \ \tr{\big(\mathscr{L}^+_t\big)^*\mathbb{J}_{\hat{\pi}_t}^{-1}\big(\hat{\pi}^{1-y'}_t\Delta \dot{\hat{H}}_t \hat{\pi}_t^{y'}\big) \mathbb{J}_{\hat{\pi}_t}(\dot{\hat{H}}^c_t) }+\mathcal{O}(\epsilon^2), \\
    \nonumber&=-\epsilon\beta^2\int^\alpha_0 dx' \int^{x'}_0 dy' \ \tr{\mathbb{J}_{\hat{\pi}_t}^{-1}\big(\hat{\pi}^{1-y'}_t\Delta \dot{\hat{H}}_t \hat{\pi}_t^{y'}\big) \mathbb{J}_{\hat{\pi}_t}\tilde{\mathscr{L}}^+_t(\dot{\hat{H}}^c_t) }+\mathcal{O}(\epsilon^2), \\
    \nonumber&=-\epsilon\beta^2\int^\alpha_0 dx' \int^{x'}_0 dy' \ \tr{\hat{\pi}^{1-y'}_t\Delta \dot{\hat{H}}_t \hat{\pi}_t^{y'}\tilde{\mathscr{L}}^+_t(\dot{\hat{H}}^c_t) }+\mathcal{O}(\epsilon^2), \\
    \nonumber&=-\epsilon\beta^2\int^\alpha_0 dx' \int^{x'}_0 dy' \ \tr{\big(\mathscr{L}^+_t\big)^*(\Delta \dot{\hat{H}}_t) \hat{\pi}_t^{y'}\dot{\hat{H}}^c_t \hat{\pi}^{1-y'}_t }+\mathcal{O}(\epsilon^2), \\
    &=-\epsilon\beta^2\int^\alpha_0 dx' \int^{x'}_0 dy' \ \tr{\big(\mathscr{L}^+_t\big)^*( \dot{\hat{H}}^c_t) \hat{\pi}_t^{y'}\dot{\hat{H}}^c_t \hat{\pi}^{1-y'}_t }+\mathcal{O}(\epsilon^2), \end{align}
where we introduced the dual Drazin inverse $\tilde{\mathscr{L}}^+_t$ from the scalar product~\eqref{eq:DB3} in the third line, hermiticity of $\mathbb{J}_{\hat{\rho}}$ and~\eqref{eq:log_exp_int} in the fourth line, in the fifth line we again used~\eqref{eq:DB3}, and in the final line we used~\eqref{eq:crossterm}. For $B(x)$ we find
\begin{align}
    \nonumber B(x)&=\epsilon^{-1}\tr{\bar{\hat{\rho}}_t^{1-x}\hat{\rho}_t^x \bigg(\mathbb{J}^{-1}_{\hat{\rho}_t}\mathscr{L}_t(\hat{\rho}_t)-\mathbb{J}^{-1}_{\hat{\rho}_t}\mathscr{L}_t(\bar{\hat{\rho}}_t) }, \\
    \nonumber&=\tr{\bar{\hat{\rho}}_t^{1-x}\hat{\rho}_t^x \bigg(\mathbb{J}^{-1}_{\hat{\rho}_t}\mathscr{L}_t(\delta\hat{\rho}_t)-\mathbb{J}^{-1}_{\hat{\rho}_t}\mathscr{L}_t(\delta\bar{\hat{\rho}}_t) }+\mathcal{O}(\epsilon^2), \\
    \nonumber&=-\beta\tr{\bar{\hat{\rho}}_t^{1-x}\hat{\rho}_t^x \dot{\hat{H}}^c_t }+\mathcal{O}(\epsilon^2), \\
    \nonumber&=\epsilon\beta^2\int^{1-x}_0 dy \tr{\dot{\hat{H}}_t^c \hat{\pi}_t^y \mathbb{J}_{\hat{\pi}_t}^{-1}\mathscr{L}^+_t\mathbb{J}_{\hat{\pi}_t}(\Delta \dot{\hat{H}}_t^d)\hat{\pi}_t^{1-y}  }+\epsilon\beta^2\int^{x}_0 dy \tr{\dot{\hat{H}}_t^c \hat{\pi}_t^{1-(x-y)} \mathbb{J}_{\hat{\pi}_t}^{-1}\mathscr{L}^+_t\mathbb{J}_{\hat{\pi}_t}(\Delta \dot{\hat{H}}_t)\hat{\pi}_t^{x-y}  }+\mathcal{O}(\epsilon^2), \\
    \nonumber&=\epsilon\beta^2\int^{x}_0 dy \tr{\dot{\hat{H}}_t^c \hat{\pi}_t^{1-(x-y)} \mathbb{J}_{\hat{\pi}_t}^{-1}\mathscr{L}^+_t\mathbb{J}_{\hat{\pi}_t}(\Delta \dot{\hat{H}}_t)\hat{\pi}_t^{x-y}  }+\mathcal{O}(\epsilon^2), \\
\end{align}
Integrating over $x$ and substituting $y'=1-(x-y)$ and $x'=x$ gives
\begin{align}
    \nonumber\int^\alpha_0 dx \ B(x)&=\epsilon\beta^2\int^\alpha_0 dx \int^x_0 dy \  \tr{\mathbb{J}_{\hat{\pi}_t}^{-1}\big(\hat{\pi}_t^{x-y}\dot{\hat{H}}_t^c \hat{\pi}_t^{1-(x-y)}\big) \mathscr{L}^+_t\mathbb{J}_{\hat{\pi}_t}(\Delta \dot{\hat{H}}_t)  }+\mathcal{O}(\epsilon^2), \\
    \nonumber&=\epsilon\beta^2\int^\alpha_0 dx' \int^{1-x}_0 dy' \  \tr{\mathbb{J}_{\hat{\pi}_t}^{-1}\big(\hat{\pi}_t^{1-y'}\dot{\hat{H}}_t^c \hat{\pi}_t^{y'}\big) \mathscr{L}^+_t\mathbb{J}_{\hat{\pi}_t}(\Delta \dot{\hat{H}}_t)  }+\mathcal{O}(\epsilon^2), \\
\end{align}
Following the same steps as with~\eqref{eq:a_part} we can further simplify this integral:
\begin{align}\label{eq:b_part}
    \int^\alpha_0 dx \ B(x)=\epsilon\beta^2\int^\alpha_0 dx' \int^{1-x'}_0 dy' \ \tr{\big(\mathscr{L}^+_t\big)^*( \dot{\hat{H}}^c_t) \hat{\pi}_t^{y'}\dot{\hat{H}}^c_t \hat{\pi}^{1-y'}_t }+\mathcal{O}(\epsilon^2),
\end{align}
We next show that the final integral vanishes:
\begin{align}
    \nonumber C(x)&=\epsilon^{-1}\int^{1-x}_0 dy \ \tr{\hat{\rho}^x_t \bigg(\text{ln} \ \hat{\rho}_t-\text{ln} \ \bar{\hat{\rho}}_t\bigg)\bar{\hat{\rho}}_t^y\mathbb{J}^{-1}_{\bar{\hat{\rho}}_t}\mathscr{L}_t(\bar{\hat{\rho}}_t)\bar{\hat{\rho}}_t^{1-x-y}}, \\
    \nonumber&=-\beta\int^{1-x}_0 dy \ \tr{\hat{\pi}^x_t \bigg(\text{ln} \ \hat{\rho}_t-\text{ln} \ \bar{\hat{\rho}}_t\bigg)\hat{\pi}_t^y\Delta \dot{\hat{H}}_t^d\hat{\pi}_t^{1-x-y}}+\mathcal{O}(\epsilon^2), \\
    \nonumber&=\beta(x-1)\tr{\Delta \dot{\hat{H}}_t^d \hat{\pi}_t \bigg(\text{ln} \ \hat{\rho}_t-\text{ln} \ \bar{\hat{\rho}}_t\bigg)}+\mathcal{O}(\epsilon^2), \\
    \nonumber &=\epsilon\beta(x-1)\tr{\Delta \dot{\hat{H}}_t^d \hat{\pi}_t \mathbb{J}^{-1}_{\hat{\pi}_t}(\delta \hat{\rho}_t-\delta\bar{\hat{\rho}}_t)}+\mathcal{O}(\epsilon^2), \\
    \nonumber \nonumber&=-\epsilon\beta(x-1)\tr{\mathbb{J}^{-1}_{\hat{\pi}_t}\mathbb{J}_{\hat{\pi}_t}(\Delta \dot{\hat{H}}_t^d) \mathscr{L}^+_t\mathbb{J}_{\hat{\pi}_t}(\dot{\hat{H}}_t^c)}+\mathcal{O}(\epsilon^2), \\
    \nonumber&=-\epsilon\beta(x-1)\tr{ \dot{\hat{H}}_t^d \mathscr{L}^+_t\mathbb{J}_{\hat{\pi}_t}(\dot{\hat{H}}_t^c)}+\mathcal{O}(\epsilon^2), \\
    &=0,
\end{align}
where we used~\eqref{eq:crossterm} in the final line. Combining~\eqref{eq:renyi2} with~\eqref{eq:a_part} and~\eqref{eq:b_part} we are left with
\begin{align}
    \frac{d}{dt}\text{ln} \ \tr{\hat{\rho}_t^\alpha \bar{\hat{\rho}}^{1-\alpha}_t(s)}\bigg|_{s=t}\simeq \epsilon\beta^2\int^\alpha_0 dx' \int^{1-x'}_x dy' \ \tr{\big(\mathscr{L}^+_t\big)^*( \dot{\hat{H}}^c_t) \hat{\pi}_t^{y'}\dot{\hat{H}}^c_t \hat{\pi}^{1-y'}_t } 
\end{align}
Setting $\alpha=1-k_B T u$ we can  obtain a relationship between the quantum covariance and quantum Renyi divergences:
\begin{align}\label{eq:renyi3}
    -u k_B T \ \dot{S}_{1-u k_B T}(\hat{\rho}_t || \bar{\hat{\rho}}_t(s))\bigg|_{s=t}\simeq-\epsilon^2\beta^2\int^\infty_0 d\nu \ \big[\hspace{-2mm}\big[ \ \text{cov}_t^{1-k_B Tu}\big(\dot{\hat{H}}^c_t(\nu),\dot{\hat{H}}_t^c\big) \ \big]\hspace{-2mm}\big].
\end{align}
Finally, we observe that the quantum CGF satisfies the symmetry $\mathcal{K}^c_q(u)=\mathcal{K}^c_q(1-u)$, which means we can integrate~\eqref{eq:renyi3} to obtain
\begin{align}
    \mathcal{K}^c_q(u)\simeq -u k_BT \int^1_0 dt \ \dot{S}_{1-u k_B T}(\hat{\rho}_t || \bar{\hat{\rho}}_t(s))\bigg|_{s=t}.
\end{align}
This concludes the proof of~\eqref{eq:CGFq_renyi} in the main text.

\section{Proof of~\eqref{eq:mono}}
\label{app:mono}
\

\noindent In this section, we will demonstrate that the cumulants of the CGF are monotonically increasing in time. To show this we will first derive an alternative expression for the CGF's in~\eqref{eq:class} and~\eqref{eq:quant}. As shown in the previous section, the conditions of detailed balance and time-translational symmetry are 
\begin{align}\label{eq:DB2}
    &\mathscr{L}_t^*-\tilde{\mathscr{L}}_t=2i \mathscr{H}_t, \\
    &[\mathscr{L}_t^*,\mathscr{H}_t]=0,
\end{align}
where $\tilde{\mathscr{L}}_t$ satisfies $\tr{\tilde{\mathscr{L}}_t(\hat{A})\hat{\pi}_t \ \hat{B}}=\tr{\hat{A} \hat{\pi}_t \mathscr{L}_t^*(\hat{B})}$ for bounded operators $\hat{A},\hat{B}$ \cite{Alicki1976}. We may express this solution as
\begin{align}
    \tilde{\mathscr{L}}_t(.)=\mathscr{L}_t\big((.)\hat{\pi}_t\big)\hat{\pi}_t^{-1}.
\end{align}
Let us now expand this superoperator in the orthonormal basis of the fixed point $\hat{\pi}_t=\sum_n p_n \ket{n}\bra{n}$:
\begin{align}\label{eq:matrix_el}
    \bra{j}\tilde{\mathscr{L}}_t(\ket{n}\bra{m})\ket{i}=\frac{p_m}{p_i}\bra{j}\mathscr{L}_t(\ket{n}\bra{m})\ket{i}.
\end{align}
We next construct another dual generator satisfying
\begin{align}
    \tr{\tilde{\mathscr{L}}_t'(\hat{A})\hat{\pi}_t^{1-x} \hat{B} \  \hat{\pi}_t^{x}}=\tr{\hat{A}  \ \hat{\pi}_t^{1-x} \mathscr{L}_t^*(\hat{B} )\hat{\pi}_t^{x}}, \ \ \ x\in\mathbb{R}.
\end{align}
The solution yields
\begin{align}
    \tilde{\mathscr{L}}_t'(.)=\hat{\pi}_t^{-x}\mathscr{L}_t\big(\hat{\pi}_t^{x}(.)\hat{\pi}_t^{1-x}\big)\hat{\pi}_t^{x-1}.
\end{align}
Expanding in the energy basis one finds
\begin{align}\label{eq:matrix_el2}
    \bra{j}\tilde{\mathscr{L}}_t'(\ket{n}\bra{m})\ket{i}=\frac{p_m}{p_i}\bigg(\frac{p_i p_n}{p_m p_j}\bigg)^x\bra{j}\mathscr{L}_t(\ket{n}\bra{m})\ket{i}.
\end{align}
We now demonstrate that~\eqref{eq:matrix_el} and~\eqref{eq:matrix_el2} are in fact equivalent following a similar method outlined in \cite{Alhambra2017}. For an operator $A$ let us consider the following decomposition in terms of modes of coherence at time $t$:
\begin{align}
    \hat{A}_\omega=\sum_{n,m}\delta\big(\omega-\log{(p_n/p_m)}\big) \ket{n}\bra{n}\hat{A}\ket{m}\bra{m},
\end{align}
It has been shown that the property of time-translational symmetry~\eqref{eq:time_cov} ensures the preservation of each mode according to~\cite{Lostaglio2015b}
\begin{align}
    \mathscr{L}_t(\hat{A}_\omega)=\big(\mathscr{L}_t(\hat{A})\big)_\omega,
\end{align}
which implies that
\begin{align}
    \log{(p_n/p_m)}\neq \log{(p_j/p_i)}\implies \bra{j}\mathscr{L}_t(\ket{n}\bra{m})\ket{i}=0.
\end{align}
Comparing~\eqref{eq:matrix_el} and~\eqref{eq:matrix_el2} we find $\bra{j}\tilde{\mathscr{L}}_t'(\ket{n}\bra{m})\ket{i}=\bra{j}\tilde{\mathscr{L}}_t(\ket{n}\bra{m})\ket{i}$, which implies
\begin{align}
    \tilde{\mathscr{L}}_t'=\tilde{\mathscr{L}}_t,
\end{align}
This means that the dual generator also fulfills the equality
\begin{align}\label{eq:DB3}
    \tr{\tilde{\mathscr{L}}_t(\hat{A})\hat{\pi}_t^{1-x} \hat{B} \  \hat{\pi}^{x}_t}=\tr{\hat{A}  \ \hat{\pi}_t^{1-x} \mathscr{L}_t^*(\hat{B})\hat{\pi}_t^{x}}, \ \ \ x\in\mathbb{R}.
\end{align}
To proceed we next introduce the following trace functional:
\begin{align}
    \langle \langle \hat{A},\hat{B} \rangle \rangle_u:=\tr{\hat{A} \ \mathbb{M}^{(u)}_{t}(\hat{B})}, 
\end{align}
where
\begin{align}
    \mathbb{M}^{(u)}_{t}(.):=-\int^u_0 dy \ \int^{1-y}_y dx \ \hat{\pi}_t^x (.)\hat{\pi}_t^{1-x},
\end{align}
Now observe that the nested commutator $\com_m[\hat{H}_t, (.)] = \mathscr{H}_t^m(.)$ (i.e.\ such that $\com_{m+1}[\hat{H}_t,(.)]=[\hat{H}_t, \com_m (\cdot)]$ with $\com_0=\id $) satisfies 
\begin{align}\label{eq:antiherm}
    \tr{\hat{A} \ \com_m[\hat{H}_t, \hat{B}]} =(-1)^{m}\tr{\com_m[\hat{H}_t, \hat{A}] \ \hat{B}}, 
\end{align}
for all $\hat{A}, \hat{B} \in \bh$ and $m\in\mathbb{Z}$. Furthermore, using $[\hat{\pi}_t,\hat{H}_t]=0$ it is straightforward to verify the commutation relations 
\begin{align}\label{eq:commC}
    [\mathbb{M}^{(u)}_{t},\com_m[\hat{H}_t, (.)]]=0, \ \ \forall m\in\mathbb{Z}.
\end{align}
Another useful property is the following symmetry of the trace-functional:
\begin{align}
    \langle \langle \hat{A},\hat{B} \rangle \rangle_u=\langle \langle \hat{B},\hat{A} \rangle \rangle_u,
\end{align}
for hermitian $\hat{A},\hat{B}$. This can be verified by using the cyclic property of the trace. Using this we find 
\begin{align}
    \nonumber\langle \langle \hat{A},\com_m[\hat{H}_t, \hat{A}] \rangle \rangle_u &=\tr{\hat{A}  \ \mathbb{M}^{(u)}_{t} \circ \com_m[\hat{H}_t, \hat{A}]}, \\
    \nonumber&=\tr{\mathbb{M}^{(u)}_{t}(\hat{A}) \com_m[\hat{H}_t, \hat{A}]}, \\
    \nonumber&=(-1)^m\tr{\com_m[\hat{H}_t, \mathbb{M}^{(u)}_{t}(\hat{A})] \ \hat{A}}, \\
    \nonumber&=(-1)^m\tr{ \mathbb{M}^{(u)}_{t} \circ \com_m[\hat{H}_t, \hat{A}] \ \hat{A}}, \\
    &=(-1)^m\langle\langle \hat{A},\com_m[\hat{H}_t, \hat{A}] \rangle\rangle_u,
\end{align}
which implies
\begin{align}\label{eq:oddterms}
    \langle\langle \hat{A},\com_{2m+1}[\hat{H}_t, \hat{A}]) \rangle\rangle_u=0, \ \forall m\in\mathbb{Z}.
\end{align}
From the detailed balance condition~\eqref{eq:DB2} and~\eqref{eq:DB3} we have
\begin{align}
    \langle\langle \hat{A}(\nu),\hat{A} \rangle\rangle_u=\langle\langle \hat{B},e^{-i\nu \mathscr{H}_t}(\hat{B})  \rangle\rangle_{u},
\end{align}
where $\nu\geq 0$ and we denote $\hat{A}(\nu)=e^{\nu \mathscr{L}_t^{*}}(\hat{A}),$ and set $\hat{B}=\hat{B}^\dagger=e^{\frac{1}{2}\nu\mathscr{L}_t^*}(\hat{A})$. Using the exponential series $e^{-i\nu \mathscr{H}_t} = \sum_{m=0}^\infty \frac{(-i \nu)^m}{ m!} \com_m[\hat{H}_t, (.)]$,  we then find
\begin{align}\label{eq:sinner}
    \nonumber\langle\langle \hat{A}(\nu),\hat{A} \rangle\rangle_u&=\langle\langle \hat{B},e^{-i
    \nu \mathscr{H}_t}(\hat{B})  \rangle\rangle_{u}, \\
    \nonumber&=\sum^\infty_{m=0}\frac{(-i\nu)^m}{m!}\langle\langle  \hat{B},\com_m[\hat{H}_t, \hat{B}]  \rangle\rangle_u, \\
    \nonumber&= \sum^\infty_{n=0}\frac{(i\nu)^{2n}}{(2n)!}\langle\langle \hat{B},\com_{2n}[\hat{H}_t, \hat{B}]  \rangle\rangle_u, \\
    \nonumber&= \sum^\infty_{n=0}\frac{(i\nu)^{2n}}{(2n)!}\tr{ \hat{B} \ \mathbb{M}^{(u)}_{t} \circ \com_{2n}[\hat{H}_t, \hat{B}]}, \\
    \nonumber&= \sum^\infty_{n=0}\frac{(i\nu)^{2n}}{(2n)!}\tr{ \hat{B} \ \com_m[\hat{H}_t, \mathbb{M}^{(u)}_{t} \circ \com_m[\hat{H}_t, \hat{B} ]]}, \\
    \nonumber&= \sum^\infty_{n=0}\frac{(-1)^n(i\nu)^{2n}}{(2n)!}\langle \langle\com_m[\hat{H}_t,\hat{B}],\com_m[\hat{H}_t,\hat{B}]  \rangle\rangle_{u}, \\
    &=\sum^\infty_{n=0}\frac{\nu^{2n}}{(2n)!}\langle\langle \com_m[\hat{H}_t,\hat{B}],\com_m[\hat{H}_t,\hat{B}]  \rangle\rangle_{u}, 
\end{align}
where we used~\eqref{eq:oddterms} in the third line,~\eqref{eq:commC} in the fifth line and~\eqref{eq:antiherm} in the penultimate line. As the final step, we use the fact that
\begin{align}\label{eq:mono3}
    (-1)^k \frac{d^k}{d u^k} \langle \langle \hat{A},\hat{A} \rangle \rangle_u \bigg|_{u=0}\geq 0  , \ \ \ \forall k\in \mathcal{Z},
\end{align}
for any $\hat{A}=\hat{A}^\dagger$, which was proven in \cite{Scandi2020}. Combining this with~\eqref{eq:sinner}, we have  
\begin{align}\label{eq:mono2}
    (-1)^k \frac{d^k}{d u^k} \langle \langle \hat{A}(\nu),\hat{A} \rangle \rangle_u \bigg|_{u=0}\geq 0  , \ \ \ \forall k\in \mathbb{Z}.
\end{align}
Finally, returning to the form of the CGF~\eqref{eq:quant}, we conclude that the cumulants are monotonically increasing with 
\begin{align}
    (-1)^k\frac{d^k}{du^k}\dot{\mathcal{K}}^c_q(u)\bigg|_{u=0}=(-1)^k\beta^2\int^\infty_0 d\nu \ \frac{d^k}{d u^k} \langle \langle \hat{H}_t^c(\nu),\hat{H}_t^c \rangle \rangle_u \bigg|_{u=0}\geq 0, \ \ \ \forall k\in \mathbb{Z},
\end{align}
which follows from~\eqref{eq:mono2}. This concludes the proof of~\eqref{eq:mono}. 

\section{The damped two-level system}
\label{app:example}

\subsection{Adiabatic Lindblad equation}

In this section, we provide details of the explicit example considered in the main text, i.e.\ a two-level system described by the Hamiltonian in Eq.~\eqref{qubit_Hamiltonian}, which we quote here again for convenience:
\begin{equation}
    \label{qubit_H}
        \hat{H}_t = \frac{\varepsilon_t}{2}\left( \cos\theta_t\hat{\sigma}_z + \sin\theta_t \hat{\sigma}_x\right).
\end{equation}
The Hamiltonian is diagonalized by the time-dependent unitary operator $\hat{U}_t = e^{-i \theta_t\hat{\sigma}_y/2}$, i.e. $U^{\dagger}_t \hat{H}_t \hat{U}_t = \tfrac{1}{2}\varepsilon_t\hat{\sigma}_z$. The time-dependent lowering operator satisfies $[\hat{H}_t,\hat{L}_t] = - \varepsilon_t\hat{L}_t$ and is given explicitly by
\begin{equation}
    \hat{L}_t = \frac{1}{2}\begin{pmatrix}-\sin\theta_t & \cos\theta_t-1 \\ \cos\theta_t+1 & \sin\theta_t \end{pmatrix}.
\end{equation}

We model the thermal bath by an infinite collection of bosonic modes coupled linearly to the system~\cite{Leggett1987}. The open-system dynamics is determined completely by the bath's inverse temperature $\beta$ and its spectral density function, assumed to take an Ohmic form $J(\omega) = \alpha \omega$ at low frequencies, with $\alpha$ a dimensionless coupling constant. Under the Born-Markov, secular and slow-driving approximations, the open-system dynamics is described by an adiabatic Lindblad master equation of the form $\dot{\hat{\rho}}_t = \mathscr{L}_t(\hat{\rho}_t)$, with the Liouvillian given by~\cite{Albash2012}
\begin{equation}\label{eq:ME}
  \mathscr{L}_t(.) = -i [\hat{H}_t,(.)] + \alpha\varepsilon_t\left(N_t+1\right)\mathscr{D}[\hat{L}_t](.) + \alpha\varepsilon_t N_t\mathscr{D}[\hat{L}_t^\dagger](.),
\end{equation}
where $N_t = \left[e^{\beta\varepsilon_t}-1\right]^{-1}$ denotes the bosonic occupation number and
\begin{equation}
  \mathscr{D}[\hat{A}](.) = \hat{A}(.)\hat{A}^\dagger - \frac{1}{2}\lbrace \hat{A}^{\dagger}\hat{A} ,(.)\rbrace,
\end{equation}
with $\lbrace\cdot,\star\rbrace$ being the anticommutator between $\cdot$ and $\star$. The bath-induced renormalisation of the qubit energy level splitting is assumed to be already incorporated into Eq.~\eqref{qubit_H}. 

The validity of the Born-Markov approximation requires that the reservoir correlation functions decay rapidly in comparison to all other timescales. The secular approximation requires that the qubit energy splitting $\varepsilon_t$ is much greater than the characteristic dissipation rate. Finally, the adiabatic master equation assumes that the Hamiltonian varies slowly in comparison to the dissipation rate. The latter can be quantified by 
\begin{equation}
    \label{characteristic_rate}
    \gamma_t = \frac{1}{2}\alpha \varepsilon_t \coth(\beta\varepsilon_t/2),
\end{equation}
which represents the average of the gain and loss rates appearing in Eq.~\eqref{eq:ME}. We therefore require the conditions $\varepsilon_t \gg \gamma_t$ and $\gamma_t\tau\gg 1$, which are well satisfied in all examples we consider. 

\subsection{Solution of the Bloch equations}

The first key quantity which we aim to compute is the full cumulant generating function
\begin{align}\label{eq:CGF2SuppMat}
    \mathcal{K}_q(u)\simeq-\epsilon \ \beta^2\int^1_0 dt  \int^\infty_0 d\nu \ \left[\hspace{-2mm}\left[\, \text{cov}^u_t\big(\dot{\hat{H}}_t(\nu),\dot{\hat{H}}_{t}\big) \right]\hspace{-2mm}\right].
\end{align}
We recall that, for convenience, Eq.~\eqref{eq:CGF2SuppMat} is expressed in terms of the rescaled time coordinate $t \in [0,1]$, normalised by the protocol time $\tau = 1/\epsilon$. In the expression above, the quantum covariances are between the power operator $\dot{\hat{H}}_{t}$ and its evolved version in the Heisenberg picture, $\dot{\hat{H}}_t(\nu) = e^{\nu \mathscr{L}^*}(\dot{\hat{H}}_t)$. The latter, in the particular model considered, is given by
\begin{equation}\label{powerop}
    \dot{\hat{H}}_t(\nu) = f_t\hat{\sigma}_z(\nu) + g_t\hat{\sigma}_x(\nu),
\end{equation}
where 
\begin{equation}
    f_t = \frac{1}{2}\left(\dot{\varepsilon}_t\cos\theta_t - \varepsilon_t\dot{\theta}_t \sin \theta_t\right) , \qquad g_t = \frac{1}{2} \left(\dot{\varepsilon}_t\sin \theta_t + \varepsilon_t \dot{\theta}_t\cos\theta_t\right),
\end{equation}
and where $\hat{\sigma}_{z,x}(\nu)$ are the evolved operators in Heisenberg picture. These satisfy $d\hat{\sigma}_{z,x}(\nu)/d\nu = \mathscr{L}^*_t\hat{\sigma}_{z,x}$, which is formally equivalent to the Bloch equations describing a damped two-level system with a fixed Lindblad generator $\mathscr{L}_t$. Introducing the Bloch vector $\mathbf{\hat{O}}(\nu) = (\hat{\sigma}_{x}(\nu),\hat{\sigma}_{y}(\nu),\hat{\sigma}_{z}(\nu))^T$, we have
\begin{equation}\label{eq:BVsystem}
 \frac{d}{d\nu}\mathbf{\hat{O}}(\nu) = \mathbf{G}_{t}\mathbf{\hat{O}}(\nu) + \mathbf{b}_{t},
\end{equation}
where the matrix $\mathbf{G}_t$ and vector $\mathbf{b}_t$ are defined by
\begin{align}
    \mathbf{G}_{t}&=
\begin{pmatrix}
\frac{1}{4} \alpha (2 N_t+1) \varepsilon_t  (\cos [2 \theta_t]-3) & \varepsilon_t  \cos [\theta_t] & -\frac{1}{4} \alpha  (2
   N_t+1) \varepsilon_t  \sin [2 \theta_t] \\
 -\varepsilon_t  \cos [\theta_t] & -\frac{1}{2} \alpha  (2 N_t+1) \varepsilon_t  & \varepsilon_t  \sin [\theta_t] \\
 -\frac{1}{4} \alpha  (2 N_t+1) \varepsilon_t  \sin [2 \theta_t ] & -\varepsilon_t  \sin [\theta_t] & -\frac{1}{4} \alpha  (2 N_t+1)
   \varepsilon_t  (\cos [2 \theta_t ]+3)
\end{pmatrix}\notag\\
\mathbf{b}_{t} &= \left(-\alpha\varepsilon_t\sin [\theta_t],  0,  -\alpha\varepsilon_t\cos [\theta_t]\right)^T.
\end{align}
Straightforward calculations finally lead to the following solution for the Bloch vector $\mathbf{\hat{O}}$
\begin{equation}\label{HeisenbergSolution}
   \begin{pmatrix}\hat{\sigma}_x(\nu)\\
   \hat{\sigma}_y(\nu)\\\hat{\sigma}_z(\nu)\end{pmatrix} = \begin{pmatrix}
   A_{xx}(\nu,t') & A_{xy}(\nu,t') & A_{xz}(\nu,t') \\
   A_{yx}(\nu,t') & A_{yy}(\nu,t') & A_{yz}(\nu,t') \\
   A_{zx}(\nu,t') & A_{zy}(\nu,t') & A_{zz}(\nu,t')
   \end{pmatrix} \begin{pmatrix}\hat{\sigma}_y(0)\\
   \hat{\sigma}_y(0)\\\hat{\sigma}_z(0)\end{pmatrix} + \begin{pmatrix}A_{x0} (\nu,t')\id\\
   A_{y0}(\nu,t')\id \\A_{z0}(\nu,t')\id \end{pmatrix},
   \end{equation}
where 
\begin{align}\label{coefficientsHeisOps}
    A_{xx}(\nu,t) &= \frac{1}{2} e^{-2 \gamma_t \nu } \left[ 2 e^{\gamma_t  \nu } \cos ^2[\theta_t] \cos [\nu \varepsilon_t ]-\cos [2 \theta_t ]+1\right]\notag\\
    A_{xy}(\nu,t) &= A_{yx}(\nu,t') = e^{-\gamma_t\nu}\cos[\theta_t] \sin [\nu  \varepsilon_t]\notag\\
    A_{xz}(\nu,t) &= A_{zx}(\nu,t') = \frac{1}{2} e^{-2 \gamma_t  \nu } \sin [2 \theta_t] \left[1-e^{\gamma_t  \nu } \cos [\nu  \varepsilon_t]\right]\notag\\
    A_{yy}(\nu,t) &= e^{-\gamma_t \nu } \cos [\nu \varepsilon_t]\notag\\
    A_{zz}(\nu,t) &= \frac{1}{2} e^{-2 \gamma_t \nu } \left[ 2 e^{\gamma_t  \nu } \sin ^2[\theta_t] \cos [\nu \varepsilon_t]+\cos [2 \theta_t ]+1\right]
\end{align}
with $\gamma_t$ given by Eq.~\eqref{characteristic_rate}. The explicit (rather cumbersome) expressions for the functions $A_{j0}(\nu,t')$ ($j=x,y,z$) will not be given here, since they drop out of the final expressions due to the fact that $[\![\mathrm{cov}^u_{t}(., \id)]\!] = 0$.

\subsection{Excess heat cumulant generating function}
\label{app:CGF_example}

We are now ready to compute the cumulant generating function of the dissipated heat in the slow-driving regime, Eq.~\eqref{eq:CGF2SuppMat}. 
As stressed above, the calculations can be simplified by first noticing that $\mathrm{cov}^u_{t}(\cdot, \id) = 0$; furthermore, it turns out that, for this model, we have that $\mathrm{cov}^u_{t}\left(\hat{\sigma}_y,\cdot\right) = 0$ and finally that $\mathrm{cov}^u_{t}\left(\hat{\sigma}_x, \hat{\sigma}_z\right) = \mathrm{cov}^u_{t}\left(\hat{\sigma}_z, \hat{\sigma}_x\right)$. Putting all this together, the result is given by
\begin{align}
    \mathcal{K}_q(u) 
    &= -\epsilon\beta^2\int_0^1  dt \, 
    \left[\!\left[\mathrm{cov}^u_{t}\left(\hat{\sigma}_z, \hat{\sigma}_z\right)\right]\!\right] \left[f^2_t \tilde{A}_{zz}(t) + f_tg_t\tilde{A}_{xz}(t)\right] \notag\\
    &\quad\quad +\left[\!\left[\mathrm{cov}^u_{t}\left(\hat{\sigma}_x, \hat{\sigma}_x\right)\right]\!\right] \left[g^2_t \tilde{A}_{xx}(t) + f_tg_t\tilde{A}_{zx}(t)\right]\notag\\
    &\quad\quad +  \left[\!\left[\mathrm{cov}^u_{t}\left(\hat{\sigma}_z, \hat{\sigma}_x\right)\right]\!\right] \left[f^2_t \tilde{A}_{zx}(t) + g^2_t \tilde{A}_{xz}(t) +  f_tg_t\left(\tilde{A}_{zz}(t) + \tilde{A}_{xx}(t)\right)\right],
\end{align}
where $\tilde{A}_{jk}(t) \equiv \int_{0}^{+\infty}d\nu A_{jk}(\nu,t)$, which can be analytically computed.
The above quantum covariances are given by
\begin{align}\label{eq:quantumcovariances}
    &\left[\!\left[\mathrm{cov}^u_{t}\left(\hat{\sigma}_x, \hat{\sigma}_x\right)\right]\!\right] = \frac{2 u \varepsilon ^2_t \sin ^2[\theta_t] (\beta -u)-2 \cos ^2[\theta_t] (-\cosh [\beta  \varepsilon_t] +\cosh [\varepsilon_t  (u-\beta )]+\cosh[ u \varepsilon_t ]-1)}{\beta
   ^2 \varepsilon_t ^2 (\cosh [\beta  \varepsilon_t ]+1)}\notag\\
   &\left[\!\left[\mathrm{cov}^u_{t}\left(\hat{\sigma}_x, \hat{\sigma}_z\right)\right]\!\right] = \frac{\sin [2 \theta_t ] \text{sech}^2\left[\frac{\beta  \varepsilon_t }{2}\right] \left(-\cosh [\beta  \varepsilon_t ]+u \varepsilon_t ^2 (\beta -u)+\cosh [\varepsilon_t
   (u-\beta )]+\cosh [u \varepsilon_t]-1\right)}{2 \beta ^2 \varepsilon_t ^2}\notag\\
   &\left[\!\left[\mathrm{cov}^u_{t}\left(\hat{\sigma}_z, \hat{\sigma}_z\right)\right]\!\right] =\frac{2 u \varepsilon_t ^2 \cos ^2[\theta_t] (\beta -u)-2 \sin ^2[\theta_t] (-\cosh [\beta  \varepsilon_t ]+\cosh [\varepsilon_t  (u-\beta )]+\cosh [u \varepsilon_t]-1)}{\beta
   ^2 \varepsilon_t ^2 (\cosh [\beta  \varepsilon_t ]+1)}.
\end{align}
Putting together all these results and substituting the functional expressions for $\varepsilon_t$ and $\theta_t$ given below in Eq.~\eqref{protocol} allows to obtain a rather cumbersome but analytic expression for the integrand of the cumulant generating function $\dot{\mathcal{K}}_q(u,t')$, where $\mathcal{K}_q(u) \equiv \epsilon\int_0^1 dt' \dot{\mathcal{K}}_q(u,t')$. The final remaining integration over the rescaled time variable $t$ however can be carried out only numerically.

\subsection{The classical part of the cumulant generating function}

Here we will provide the details of the calculations of the classical part of the cumulant generating function, in order to single out the quantum contributions due to coherences. Following the main theory outlined above and in the main text, we first need to compute the diagonal part of the power operator onto the instantaneous energy eigenbasis, i.e. \begin{equation}
    \dot{\hat{H}}^d_{t} = \mathcal{D}_{\hat{H}_t}\left(\dot{\hat{H}}_t\right) = \frac{1}{2}\dot{\varepsilon}_t \left(\cos[\theta_t]\hat{\sigma}_z + \sin[\theta_t]\hat{\sigma}_x\right).
\end{equation}
In order to then find $\dot{\hat{H}}_{t}^d(\nu)$ we can exploit the solutions Eq.~\eqref{HeisenbergSolution} for the evolved operators $\hat{\sigma}_z(\nu),\, \hat{\sigma}_x(\nu)$.
What thus remains to be calculated in order to determine $\mathcal{K}^d_q(u)$ defined in Eq.~\eqref{eq:class} is the symmetrised covariance $\text{cov}_{t}(\dot{\hat{H}}_{t}^d(\nu),\dot{\hat{H}}_t^d) = \frac{1}{2}\tr{\lbrace\dot{\hat{H}}_t^d(\nu),\dot{\hat{H}}_{t}^d\rbrace\hat{\pi}_{t}}-\tr{\dot{\hat{H}}_{t}^d(\nu)\hat{\pi}_{t}}\tr{\dot{\hat{H}}_{t}^d\hat{\pi}_{t}}$.
The calculations are simplified first by noticing that this quantity is symmetric in its two arguments and furthermore that $\text{cov}_{t}(\cdot,\id) = 0$ and $\text{cov}_{t}(\cdot,\hat{\sigma}_y) = 0$. The remaining symmetric covariances are given by
\begin{align}
  \text{cov}_{t}(\hat{\sigma}_x,\hat{\sigma}_x) &=   \frac{\left(e^{\beta \varepsilon_t}-1\right)^2 \cos [2 \theta_t]+6 e^{\beta \varepsilon_t}+e^{2 \beta \varepsilon_t}+1}{2 \left(e^{\beta \varepsilon_t }+1\right)^2}\\
  \text{cov}_{t}(\hat{\sigma}_x,\hat{\sigma}_z) &=
 -\sin[\theta_t]\cos[\theta_t] \tanh ^2\left(\frac{\beta \varepsilon_t }{2}\right)\\
  \text{cov}_{t}(\hat{\sigma}_z,\hat{\sigma}_z) &= 1-\cos ^2[\theta_t] \tanh ^2\left(\frac{\beta \varepsilon_t }{2}\right).
\end{align}
A straightforward calculation finally shows that
\begin{align}
    \mathcal{K}^d_q(u) &= -\epsilon (u^2-\beta u)\int_0^1 dt' \int_0^{\infty}d\nu \mathrm{cov}_{t}\left(\dot{\hat{H}}^d_{t}(\nu),\dot{\hat{H}}^d_{t}\right)\notag \\
    &= -\epsilon (u^2-\beta u)\int_0^1 dt'\left(\frac{1}{2}\dot{\varepsilon}_t\right)^2\left[ \mathrm{cov}_{t}\left(\hat{\sigma}_z, \hat{\sigma}_z\right)\cos[\theta_t]\left(\cos[\theta_t]\tilde{A}_{zz}(t) + \sin[\theta_t]\tilde{A}_{xz}(t)\right) \right. \notag \\ & \qquad 
  \qquad \qquad 
  \qquad\qquad 
  \qquad+ \left. \mathrm{cov}_{t}\left(\hat{\sigma}_x, \hat{\sigma}_x\right)\sin[\theta_t]\left(\sin[\theta_t]\tilde{A}_{xx}(t) + \cos[\theta_t]\tilde{A}_{zx}(t)\right)\right.\notag\\
    &\qquad\qquad \qquad 
  \qquad \qquad 
  \qquad+ \left.\mathrm{cov}_{t}\left(\hat{\sigma}_x, \sigma_z\right)\left(\tilde{A}_{zx}(t) +\sin[\theta_t]\cos[\theta_t]\left(\tilde{A}_{xx}(t) + \tilde{A}_{zz}(t)\right) \right)\right].
\end{align}



\section{Monte Carlo simulations and results}
\label{app:trajectories}

\begin{figure}
    \centering
    \includegraphics[width=0.5\linewidth, trim = 0mm 80mm 0mm 80mm, clip] {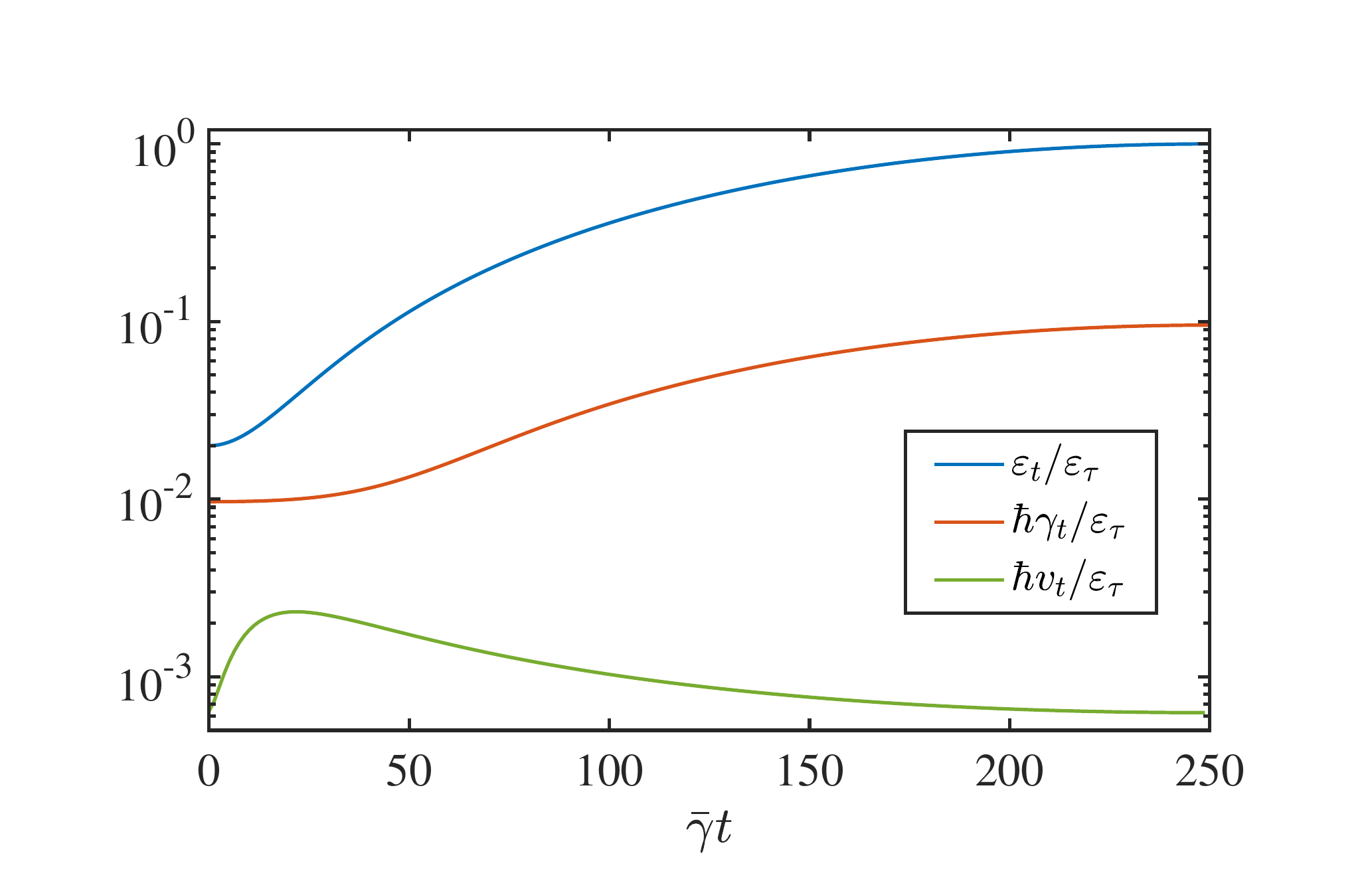}
    \caption{Comparison between the energy splitting $\varepsilon_t$, the dissipation rate $\gamma_t = \alpha\varepsilon_t \coth(\beta\varepsilon_t/2)/2\hbar$ and the instantaneous driving speed $v_t = \sqrt{(\dot{\varepsilon}_t/\varepsilon_t)^2 + \dot{\theta}^2}$. The conditions for the secular and slow-driving approximations are seen to hold at all times. The parameters are the same as in Fig.~\ref{fig:distribution_outliers}, i.e.\ $\alpha = 0.191$, $\varepsilon_0 = 0.02\varepsilon_\tau$, $\beta\varepsilon_\tau = 20$ and $\bar{\gamma}\tau=250$.}
    \label{fig:scales}
\end{figure}

We now give further details on the Monte Carlo trajectory simulations used to obtain Fig.~\ref{fig:distribution_outliers}. In the following discussion we restore the non-rescaled time coordinate so that the protocol takes place in the time interval $t\in [0,\tau]$. We follow the standard quantum-jump approach~\cite{Plenio1998} to unravel the Lindblad equation defined by Eq.~\eqref{eq:ME} into pure-state trajectories. The initial state $\ket{\psi_0}$ is randomly selected from the eigenstates of $\hat{H}_0$ according to the corresponding Boltzmann distribution. The time at which the first jump occurs is chosen by sampling from the waiting-time distribution $w_t=-(d/dt)\langle\psi_t|\psi_t\rangle$, where $\ket{\psi_t}$ obeys the non-Hermitian Schr\"odinger equation $(d/dt)\ket{\psi_t} = -i \hat{H}^{\rm eff}_t \ket{\psi_t}$ with initial condition $\ket{\psi_0}$, and the effective Hamiltonian is
\begin{equation}
    \label{non_hermitian_Hamiltonian}
    \hat{H}^{\rm eff}_t = \hat{H}_t - \frac{i\alpha \varepsilon_t}{2}\left( (N_t+1)\hat{L}^\dagger_t \hat{L}_t + N_t \hat{L}_t\hat{L}_t^\dagger\right).
\end{equation}
This evolution equation is solved efficiently using a fourth-order Runge-Kutta algorithm with adaptive time-step control~\cite{Press2007}. If a jump occurs at time $t$, it corresponds either to emission, $\ket{\psi_t}\to \hat{L}_t\ket{\psi_t}$, or absorption,  $\ket{\psi_t}\to \hat{L}^\dagger_t\ket{\psi_t}$, with probabilities given by the excited or ground state populations, respectively. The normalised post-jump state is then taken as a new initial condition and the procedure is repeated until the final time is reached. The heat distribution is constructed by simulating many such trajectories and recording the total quantity of energy transferred to the environment during each one, as in Eq.~\eqref{eq:heat}.

The specific protocol we consider is defined by
\begin{equation}
\label{protocol}
{\varepsilon_t = \varepsilon_0 + (\varepsilon_\tau- \varepsilon_0)\sin^2(\pi t/2\tau)}, \qquad {\theta_t = \pi(t/\tau-1)}.
\end{equation}
Erasure corresponds to choosing the initial spectral gap to be far below the thermal energy, $\varepsilon_0 \ll k_B T$, and the final spectral gap to greatly exceed this energy, $\varepsilon_\tau \gg k_BT$. This ensures that the respective thermal states are effectively given by $\hat{\pi}_0 \simeq \id/2$ and $\hat{\pi}_\tau\simeq \ket{0}\bra{0}$, respectively. The slow-driving regime corresponds to the case where the Hamiltonian changes slowly in comparison to the characteristic relaxation rate $\gamma_t = \tfrac{1}{2}\alpha \varepsilon_t \coth(\beta\varepsilon_t/2)$. The instantaneous driving speed can be quantified by the parameter $v_t = \sqrt{(\dot{\varepsilon}_t/\varepsilon_t)^2 + \dot{\theta}^2}$, which takes into account the rate of change of both the energy splitting $\varepsilon_t$ and the mixing angle $\theta_t$. In Fig.~\ref{fig:scales} we show that the conditions for the validity of both the slow-driving  and secular approximations, namely $\hbar v_t \ll \hbar \gamma_t \ll \varepsilon_t$, hold at all times during the evolution for the parameters considered in this work. 

\begin{figure}
    \centering
    \includegraphics[width=0.5\linewidth, trim = 0mm 80mm 0mm 80mm, clip]{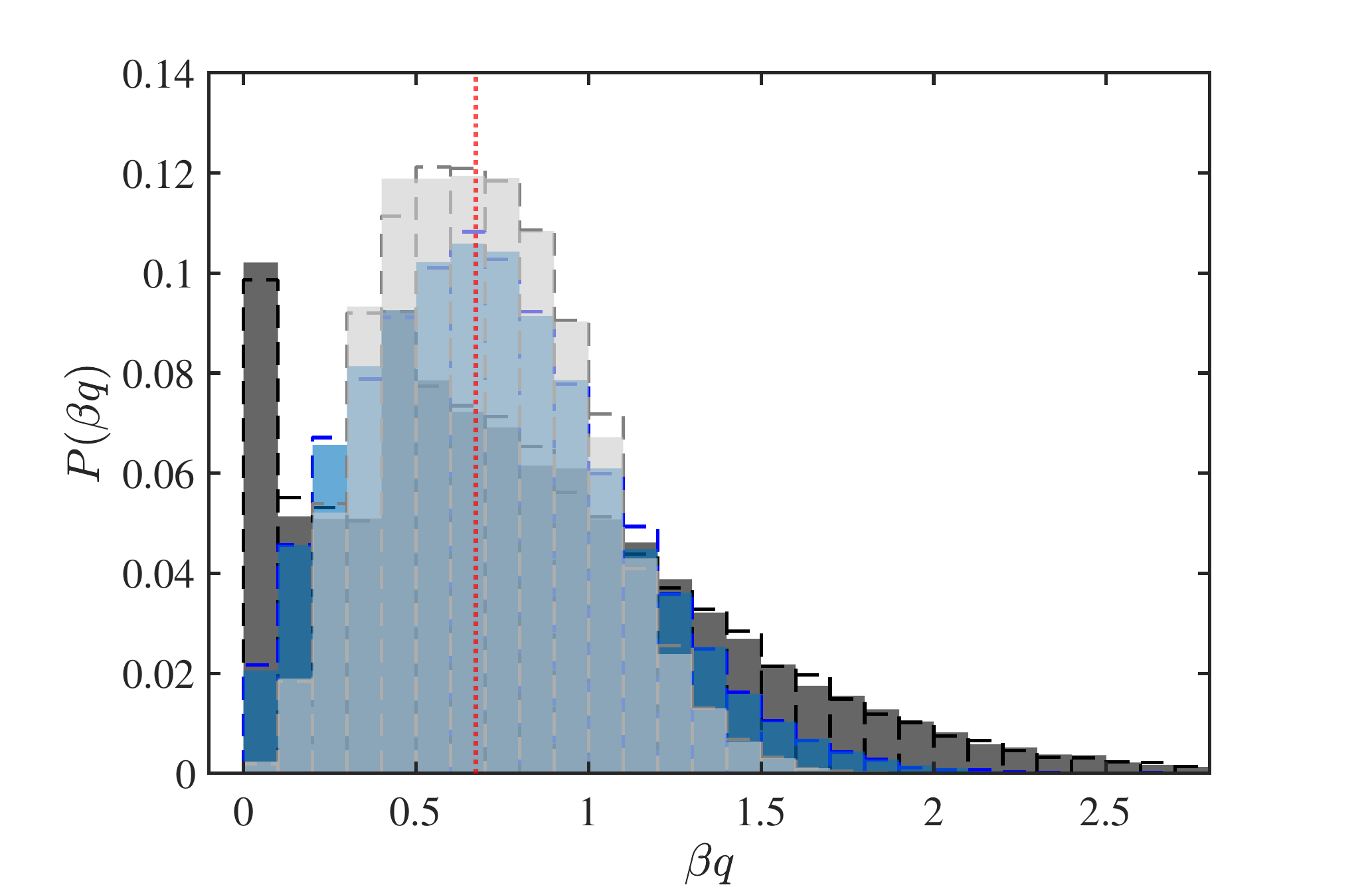}
    \caption{Probability distribution of heat transferred during slow erasure protocols. The data correspond to classical (dashed lines) and quantum (solid bars) protocols of duration $\bar{\gamma}\tau = 100$ (dark grey), $\bar{\gamma}\tau = 250$ (blue) and $\bar{\gamma}\tau = 500$ (light grey). The Landauer limit $ \beta q=-\Delta S$ is shown by the red dotted line. Other parameters are the same as in Fig.~\ref{fig:distribution_outliers}, i.e.\ $\alpha = 0.191$, $\varepsilon_0 = 0.02\varepsilon_\tau$, $\beta\varepsilon_\tau = 20$.}
    \label{fig:quantum_classical_distributions}
\end{figure}

In Fig.~\ref{fig:quantum_classical_distributions} we show the heat distributions obtained for various different values of $\tau$, considering both quantum (where $\theta_t$ is given by Eq.~\eqref{protocol}) and classical (with $\theta_t=0$) protocols. We find that the bulk of the distribution is very similar between the quantum and corresponding classical protocols. The only qualitative difference is seen in the extreme outliers which do not occur in classical protocols. Such outliers may correspond either to large heat transfer, $q\gg k_BT$, or, less frequently, a negative total heat transfer $q<0$, which cannot occur classically. In both the classical and quantum case, the bulk of the distribution becomes increasingly concentrated around the Landauer bound and converges to a Gaussian shape as $\tau$ is increased. 

\end{document}